\documentclass[12pt]{article}
\usepackage{amsmath, amssymb}
\usepackage{graphicx}
\usepackage{natbib}
\usepackage{url} 
\usepackage{color}
\usepackage{geometry}
\usepackage{setspace}

\begin{document}
		
\title{\bf Laplace approximation for fast Bayesian inference in generalized additive models based on penalized regression splines}
\author{Oswaldo Gressani $^{a,}$\thanks{Corresponding author. E-mail address: \textit{$oswaldo\_gressani@hotmail.fr$}} \  and Philippe Lambert  $^{a,b}$\hspace{.2cm}\\
\\
$^a$ Institute of Statistics, Biostatistics and Actuarial Sciences (ISBA),\\  Universit\'e catholique de Louvain, Voie du Roman Pays 20, \\ B-1348, Louvain-la-Neuve, Belgium\\
\\
$^b$ Institut de Recherche en Sciences Sociales (IRSS), \\
M\'ethodes Quantitatives en sciences sociales, \\
Universit\'e de Li\`ege, Place des Orateurs 3,\\
B-4000, Li\`ege, Belgium
\date{}
}

\maketitle

\vspace{-1cm}

\begin{abstract}
Generalized additive models (GAMs) are a well-established statistical tool for modeling complex nonlinear relationships between covariates and a response assumed to have a conditional distribution in the exponential family. In this article, P-splines and the Laplace approximation are coupled for flexible and fast approximate Bayesian inference in GAMs. The proposed Laplace-P-spline model contributes to the development of a new methodology to explore the posterior penalty space by considering a deterministic grid-based strategy or a Markov chain sampler, depending on the number of smooth additive terms in the predictor. Our approach has the merit of relying on closed form analytical expressions for the gradient and Hessian of the approximate posterior penalty vector, which enables to construct accurate posterior pointwise and credible set estimators for latent field variables at a relatively low computational budget even for a large number of smooth additive components. Based upon simple Gaussian approximations of the conditional latent field posterior, the suggested methodology enjoys excellent statistical properties. The performance of the Laplace-P-spline model is confirmed through different simulation scenarios and the method is illustrated on two real datasets.
\end{abstract}


\newpage

\setstretch{1}

\section{Introduction}
\label{sec:intro}

\indent Generalized additive models (GAMs) \citep{hastie1986generalized,hastie1987generalized} extend generalized linear models \citep{nelder1972generalized} by having nonlinear smooth functions of quantitative covariates entering the linear predictor: they enable to relate in a flexible way covariates to the mean of a conditional distribution in the exponential family. The monograph of \cite{hastie1990generalized} gives a thorough introduction to additive regression structures and largely contributed to the dissemination of this model class. The textbooks by \cite{ruppert2003semiparametric} and \cite{wood2017generalized} provide a complete and comprehensive treatment of GAMs, emphasizing on semiparametric methods and penalized regression splines.\\
\indent There exists a large variety of regression splines in the literature for modeling the smooth terms in a GAM, for instance P-splines \citep{eilers1996flexible}, thin plate splines \citep{wood2003thin}, O'Sullivan penalized splines \citep{wand2008semiparametric} or adaptive splines \citep{krivobokova2008fast} to cite the most popular instances. This article focuses exclusively on P-spline smoothers for two main reasons. First, the penalty matrix can be effortlessly constructed from basic difference formulas, keeping the penalization scheme simple and the P-spline approach numerically stable. Second, the attractiveness of P-splines lies in its rather natural extension to a Bayesian setting \citep{lang2004bayesian} and from the efficiency of working with sparse bases and penalties for sampling-free approximate Bayesian inference or Markov chain Monte Carlo (MCMC) methods.\\
\indent As MCMC techniques can be subject to poor chain convergence and tend to carry a heavy computational burden, \cite{rue2009approximate} introduced an approximate Bayesian methodology based on Laplace approximations termed Integrated Nested Laplace Approximations (INLA), a completely sampling-free framework that delivers accurate and fast approximations of posterior marginals in structured additive regression models. More recent articles on fast approximate likelihood or Bayesian-based inference include \cite{luts2014real}, \cite{wand2017fast} and \cite{hui2019semiparametric} among others. Although INLA is a well-tailored approach for making inference in a variety of statistical models, there is room for further computational improvements when considering the specific class of GAMs. In particular, the use of numerical differentiation techniques in INLA to obtain finite difference approximations to the gradient and Hessian matrix of the posterior penalty vector can be replaced by their exact analytical expressions, yielding more efficient algorithms for model fitting. Furthermore, as the computational cost grows exponentially with the dimension of the penalty vector, in grid-based derivation of the marginal posterior of the regression parameters, alternative strategies are required to explore the posterior penalty space when the number of additive terms is large.\\
\indent Taken separately, P-splines and INLA have made an impressive impact in the statistical community and initiated a flourishing literature in diversified domains (see e.g. \citealp{eilers2015twenty}; \citealp{rue2017bayesian}), yet few references attempted to unify the strength of both approaches. In the present article, we borrow some ideas from INLA and combine them with P-splines to design the Laplace-P-spline (LPS) methodology, a novel unified approach for approximate Bayesian inference in generalized additive models. Our methodology is free of the numerical differentiation scheme found in INLA, as it relies on closed analytical expressions for the gradient and Hessian required during computation. It enables not only to fasten our code, but also offers a clear insight on the equations governing the implementation of the model. Moreover, we exploit this analytical availability to develop a novel cost-effective grid exploration algorithm to explore the posterior of the hyperparameters corresponding, in our specific context, to the penalty parameters controlling the smoothness of each additive term. The method accounts for possible asymmetries in the posterior hyperparameter space by applying a moment-matching technique with reference to the skew-normal family. Finally, in response to the ``curse of dimensionality'' related to the increase in computational resources with the hyperparameter dimension, we suggest to embed a regular MCMC algorithm to explore the hyperparameter posterior instead of the classic grid exploration when the dimension grows above a certain threshold. The latter idea of combining Laplace approximations with MCMC methods can be found in \cite{yoon2011inference} and more recently in \cite{gomez2017markov}. \\
\indent The remainder of the article is outlined as follows. In Section 2 the Bayesian Laplace-P-spline generalized additive model is formulated and the Laplace approximation to the conditional posterior of latent field variables is derived. To efficiently explore the approximate marginal posterior of the penalty parameters, we propose a strategy that alternates between a deterministic grid and an independence Metropolis-Hastings sampler depending on the number of smooth additive components. The chosen penalty values are then used to approximate the marginal posterior for latent field variables along with their associated pointwise credible intervals. A detailed simulation study is presented in Section 3 together with comparisons against a popular benchmark method. Section 4 illustrates the Laplace-P-spline model on two real datasets and Section 5 closes the paper with concluding remarks and sketches future research prospects.

\section{The Laplace-P-spline generalized additive model}

\subsection{Flexible modeling with P-splines}

\indent We consider a GAM where the response variable has a distribution belonging to the one-parameter exponential family $y_i \sim \text{EF}(\gamma_i,\varkappa)$ characterized by densities of the form:

\vspace{-0.3cm}

\begin{eqnarray}
p(y_i; \gamma_i, \varkappa)=\exp\left( \frac{y_i \gamma_i-s(\gamma_i)}{\varkappa}+c(y_i, \varkappa)\right), \label{exponential_family}
\end{eqnarray}

\vspace{0.1cm}

\noindent where $s(\cdot)$ is a twice continuously differentiable real-valued function and $c(\cdot,\cdot)$ another real function, $\varkappa>0$ is a known scale or dispersion parameter and $\gamma_i$ is the natural or canonical parameter. Using well-known properties of the score function (\citealp{mccullagh1989generalized}), one can show that the mean and variance of the response are $\mathbb{E}(y_i):=\mu_i=s'(\gamma_i)$ and $\text{Var}(y_i)=\varkappa s''(\gamma_i)$ respectively. Let $\mathcal{D}=\{(y_i, \mathbf{x}_i, \mathbf{z}_i)_{i=1}^n\}$ be a sample of $n$ independent observations, where $\mathbf{x}_i=(x_{i1},\dots,x_{iq})^T$ is a vector of continuous covariates and $\mathbf{z}_i=(z_{i1},\dots,z_{ip})^T$ a vector of additional covariates (possibly categorical). The link function $g(\cdot)$ relates the mean response to the additive predictor as follows:

\vspace{-0.5cm}

\begin{eqnarray}
g(\mu_i):=\varrho_i=\beta_0+\beta_1 z_{i1}+\dots+\beta_p z_{ip}+f_1(x_{i1})+\dots+f_q(x_{iq}),\ \ i=1,\dots,n.  \label{generalized_additive_model} 
\end{eqnarray}

\noindent In the spirit of the P-spline approach proposed in \cite{eilers1996flexible}, the unknown smooth functions $f_j,\ j=1,\dots ,q$ are modeled with rich cubic B-spline bases and a discrete penalty on neighboring spline coefficients is imposed for controlling the roughness of the fit. Mathematically:

\vspace{-1cm}

\begin{eqnarray}
f_j(x_{ij})=\sum_{k=1}^K \theta_{jk} b_{jk}(x_{ij}), \ \ j=1,\dots,q,
\label{Bspline_spec}
\end{eqnarray}

\noindent where for simplicity the same number $K$ of basis functions $b_{jk}(\cdot)$ is assumed for every $f_j$. The vector of B-spline coefficients associated to function $f_j$ is  $\boldsymbol{\theta}_j=(\theta_{j1},\dots,\theta_{jK})^T$, while the collection of all spline coefficients present in the model is  $\boldsymbol{\theta}=(\boldsymbol{\theta}_1^T,\dots,\boldsymbol{\theta}_q^T)^T$ and the vector of B-spline functions at $x_{ij}$ is written as  ${\bf{b}}_j(x_{ij})=(b_{j1}(x_{ij}),\dots,b_{jK}(x_{ij}))^T$. Model flexibility is compensated by a roughness penalty on finite differences of the coefficients of contiguous B-splines, $\boldsymbol{\theta}^T \mathcal{P}(\boldsymbol{\lambda}) \boldsymbol{\theta}$, with block diagonal matrix $\mathcal{P}(\boldsymbol{\lambda})$ expressed compactly using a Kronecker product:

\vspace{-0.3cm}

\begin{eqnarray}
\mathcal{P}(\boldsymbol{\lambda}):=\text{diag}(\lambda_1,\dots,\lambda_q) \otimes P=\begin{pmatrix}
\lambda_1P & 0 & \dots & 0\\
0 & \lambda_2P& \dots & 0 \\
\vdots & \dots & \ddots & 0 \\
0 & \dots & 0 & \lambda_qP
\end{pmatrix}, \nonumber 
\end{eqnarray}

\vspace{0.2cm}

\noindent where $\boldsymbol{\lambda}=(\lambda_1,\dots,\lambda_q)^T$ is a vector of positive penalty parameters and $P=D_r^T D_r+\epsilon I_K$ is a penalty matrix resulting from the product of $r${th} order difference matrices $D_r$ of dimension $(K-r) \times K$ to which a diagonal perturbation $\epsilon I_K$ is added (with $\epsilon=10^{-6}$, say), so that $P$ is full rank. From a Bayesian perspective, \cite{lang2004bayesian} suggest to obtain the roughness penalty by imposing a multivariate Gaussian prior on the spline amplitudes $\boldsymbol{\theta}\vert \boldsymbol{\lambda} \sim\mathcal{N}_{\text{dim}(\boldsymbol{\theta})}\big(0, \mathcal{P}^{-1}(\boldsymbol{\lambda})\big)$. Furthermore, a Gaussian prior is assumed on the regression coefficients  $\boldsymbol{\beta}=(\beta_0,\dots,\beta_p)^T$, more specifically  $\boldsymbol{\beta} \sim \mathcal{N}_{\text{dim}(\boldsymbol{\beta})}(0,  V^{-1}_{\boldsymbol{\beta}})$ with matrix $V_{\boldsymbol{\beta}}=\zeta I_{p+1}$ and small precision (say $\zeta=10^{-5}$). The latent field of the model is written as  $\boldsymbol{\xi}=(\boldsymbol{\beta}^T,\boldsymbol{\theta}^T)^T$ and includes the regression and spline coefficients with prior distribution $\boldsymbol{\xi} \vert \boldsymbol{\lambda} \sim \mathcal{N}_{\text{dim}(\boldsymbol{\xi})}\big(0, \big(Q^{\boldsymbol{\lambda}}_{\boldsymbol{\xi}}\big)^{-1}\big)$ and precision matrix:

\vspace{0.1cm}

\[Q^{\boldsymbol{\lambda}}_{\boldsymbol{\xi}}:=Q_{\boldsymbol{\xi}}(\boldsymbol{\lambda})=\begin{pmatrix}
V_{\boldsymbol{\beta}} & 0  \\
0 & \mathcal{P}(\boldsymbol{\lambda})  
\end{pmatrix}.\]

\vspace{0.3cm}

\noindent Without loss of generality, the covariates $\mathbf{z}_i$ are centered around their mean value. Let $\bar{z}_l=n^{-1}\sum_{i=1}^n z_{il},\ l=1,\dots,p$ and write the centered design matrix Z and B-spline matrices $B_j$ for $j=1,\dots,q$ as follows:

\vspace{0.2cm}

\[Z=\begin{bmatrix}
1 & (z_{11}-\bar{z}_1) & \dots & (z_{1p}-\bar{z}_p) \\
\vdots & \vdots & \vdots & \vdots \\
1 & (z_{n1}-\bar{z}_1) & \dots & (z_{np}-\bar{z}_p) 
\end{bmatrix}, \ \ 
B_j=\begin{bmatrix}
b_{j1}(x_{1j}) & \dots & b_{jK}(x_{1j}) \\
\vdots & \vdots & \vdots \\
b_{j1}(x_{nj}) & \dots & b_{jK}(x_{nj})
\end{bmatrix}.\] 

\vspace{0.1cm}

\subsection{Identifiability and priors}

To reach an identifiable model, we impose the following centering on the B-spline matrices $\tilde{B}_j=B_j-(\boldsymbol{1}_n \boldsymbol{1}_L^T/L)\breve{B}_j,\ j=1,\dots,q$, where $\boldsymbol{1}_n$ and $\boldsymbol{1}_L$ are column vector of ones of length $n$ and $L$ respectively and $\breve{B}_j$ is a B-spline matrix computed on a fine grid of equidistant values on the domain of $f_j$. This identifiability constraint centers the additive functional components around their average value. To ensure that all spline coefficients can be estimated in a unique way, we follow \cite{wood2017generalized} and fix the $K${th} element of each spline vector $\boldsymbol{\theta}_j$ to zero and delete the $K${th} column in $\tilde{B}_j$ and difference matrix $D_r$. Hence $\tilde{B}_j$ has $K-1$ columns and the latent vector has dimension $\text{dim}(\boldsymbol{\xi})= q \times (K-1)+p+1$. This is to be contrasted with the model setting in INLA, where the latent field dimension grows with sample size $n$.\\
\indent Following \cite{jullion2007robust}, robust priors are specified on the roughness penalty parameters with a conjugate Gamma family having a hierarchical structure $\lambda_j \vert \delta_j \sim \mathcal{G}(\nu/2,(\nu \delta_j)/2),$ $j=1,\dots,q$. An uninformative distribution is imposed on the hyperparameter $\delta_j \sim \mathcal{G}(a_{\delta},b_{\delta}),\  j=1,\dots,q$ with mean $a_{\delta}/b_{\delta}$ and variance $a_{\delta}/b_{\delta}^2$. The authors show that when $a_{\delta}=b_{\delta}$ are calibrated to a small value (say $10^{-4}$), the fitted curves are not sensitive to the value taken by $\nu$ (here  $\nu=3$). The penalty parameters are gathered in the vector $\boldsymbol{\eta}=(\boldsymbol{\lambda}^T, \boldsymbol{\delta}^T)^T$. Taking into account the identifiability constraint, the additive predictor in (\ref{generalized_additive_model}) can be expressed compactly as $\boldsymbol{\varrho}=B\boldsymbol{\xi}$, where $B$ is a side by side configuration of design matrices, $B=[Z: \tilde{B}_1:\dots: \tilde{B}_q]$ and corresponds to the full design matrix of the model. The Bayesian model is summarized as follows:

\vspace{-0.2cm}

\begin{eqnarray}
&& y_i\vert \boldsymbol{\xi} \sim  \text{EF}(\gamma_i,\varkappa), \ \ i=1,\dots,n, \nonumber \\
&&\boldsymbol{\theta}\vert \boldsymbol{\lambda} \sim\mathcal{N}_{\text{dim}(\boldsymbol{\theta})}\big(0,  \mathcal{P}^{-1}(\boldsymbol{\lambda})\big), \nonumber \\
&& \boldsymbol{\xi} \vert \boldsymbol{\lambda} \sim \mathcal{N}_{\text{dim}(\boldsymbol{\xi})}\big(0, \big(Q^{\boldsymbol{\lambda}}_{\boldsymbol{\xi}}\big)^{-1}\big), \nonumber \\
&& \lambda_j \vert \delta_j \sim \mathcal{G}(\nu/2,(\nu \delta_j)/2), \ \ j=1,\dots,q, \nonumber \\
&& \delta_j \sim \mathcal{G}(a_{\delta},b_{\delta}), \nonumber \ \ j=1,\dots,q.
\end{eqnarray}

\vspace{-0.1cm}

\subsection{Approximated conditional latent field posterior}

Let us denote by $\ell(\boldsymbol{\xi};\mathcal{D})=(1/\varkappa) \sum_{i=1}^n \left(y_i \gamma_i-s(\gamma_i) \right)+c$, with $c:=\sum_{i=1}^n c(y_i,\varkappa)$ (for ease of notation) the log-likelihood function following from the set-up described at the beginning of Section 2.1. From the standard theory of exponential families, we know that the score vector is given by 
$\nabla_{\boldsymbol{\xi}} \ell(\boldsymbol{\xi}; \mathcal{D})=B^TW D_g (\boldsymbol{y}-\boldsymbol{\mu})$, where $W:=\text{diag}(w_1,\dots,w_n)$ is a diagonal matrix with weights on the diagonal defined as $w_i:=\left(\text{Var}(y_i) [g'(\mu_i)]^2\right)^{-1}$ and $D_g=\text{diag}(g'(\mu_1),\dots, g'(\mu_n))$. Moreover, the observed Fisher information matrix (equal to the negative Hessian of the log-likelihood) is given by 
$-\nabla_{\boldsymbol{\xi}}^2 \ell(\boldsymbol{\xi}; \mathcal{D}) =B^TWB$. Using Bayes' theorem, the conditional posterior of the latent field is proportional to the product of the likelihood and prior, which can be written as 
$p(\boldsymbol{\xi} \vert \boldsymbol{\lambda}, \mathcal{D}) \propto \exp\left( \ell(\boldsymbol{\xi}; \mathcal{D})-(1/2) \boldsymbol{\xi}^T Q^{\boldsymbol{\lambda}}_{\boldsymbol{\xi}} \boldsymbol{\xi}\right)$. Using the Newton-Raphson algorithm, we compute the mode $\hat{\boldsymbol{\xi}}_{\boldsymbol{\lambda}}$ of the conditional posterior $p(\boldsymbol{\xi} \vert \boldsymbol{\lambda}, \mathcal{D})$ and use Laplace's method to approximate the latter by a normal density denoted by $\tilde{p}_G(\boldsymbol{\xi} \vert \boldsymbol{\lambda}, \mathcal{D})$. After convergence of the iterative algorithm, we recover a Gaussian centered around $\hat{\boldsymbol{\xi}}_{\boldsymbol{\lambda}}=(B^T \widetilde{W}B+Q^{\boldsymbol{\lambda}}_{\boldsymbol{\xi}})^{-1} \widetilde{\boldsymbol{\varpi}}$ with variance-covariance matrix equal to the inverse of the sum of the negative Hessian of the log-likelihood and the precision matrix $Q^{\boldsymbol{\lambda}}_{\boldsymbol{\xi}}$, i.e. $\widehat{\Sigma}_{\boldsymbol{\lambda}}=(B^T \widetilde{W}B+Q^{\boldsymbol{\lambda}}_{\boldsymbol{\xi}})^{-1}$, where $\widetilde{W}$ is the weight matrix at convergence and $\widetilde{\boldsymbol{\varpi}}$ is the vector at convergence that results from the sequence $\boldsymbol{\varpi}^{(0)},\boldsymbol{\varpi}^{(1)},\boldsymbol{\varpi}^{(2)},\dots,$ with $\boldsymbol{\varpi}^{(0)}:=(1/\varkappa) B^T\big(\boldsymbol{y}-\boldsymbol{\mu}\big(\boldsymbol{\xi}^{(0)}\big)\big)+B^TW\big(\boldsymbol{\xi}^{(0)}\big)B\boldsymbol{\xi}^{(0)}$ computed from an initial guess $\boldsymbol{\xi}^{(0)}$ of the latent field vector. The Laplace approximation $\tilde{p}_G(\boldsymbol{\xi} \vert \boldsymbol{\lambda}, \mathcal{D})$ will be used to approximate the integrand entering the computation of the marginal posterior for $\boldsymbol{\xi}$:

\vspace{-0.3cm}

\begin{eqnarray} \label{integrate_out}
p(\boldsymbol{\xi} \vert \mathcal{D})=\int_{\mathbb{R}_{++}^q} p(\boldsymbol{\xi} \vert \boldsymbol{\lambda},\mathcal{D})\ p(\boldsymbol{\lambda} \vert \mathcal{D})\ d\boldsymbol{\lambda}.  
\end{eqnarray}

\noindent Quadrature points to compute \eqref{integrate_out} will be obtained in the next section using an approximation of the marginal posterior $p(\boldsymbol{\lambda} \vert \mathcal{D})$ for the vector of penalty parameters.

\subsection{Marginal posterior of the penalty parameters}

An indispensable intermediate step to reach an approximated version for the marginal posterior of the regression and spline variables $\boldsymbol{\xi}$
is to obtain the marginal posterior of the vector $\boldsymbol{\lambda}$ of penalty parameters. In that endeavor, we first derive an approximation of $p(\boldsymbol{\eta} \vert \mathcal{D})$ in the philosophy of \cite{leonard1982comment}, \cite{tierney1986accurate} and \cite{rue2009approximate} and show how $\boldsymbol{\delta}$ can be integrated out, resulting in an approximation of the marginal posterior for the roughness penalty vector $\boldsymbol{\lambda}$. The gradient and Hessian of that log posterior are analytically derived and will prove to be very useful to explore the support of the posterior distribution of the penalty vector.

\subsubsection{Approximation to the marginal posterior of the penalty parameters}

\noindent The posterior of the hyperparameter vector is given by:

\vspace{-0.5cm}

\begin{eqnarray}
p(\boldsymbol{\eta} \vert \mathcal{D})&=&\frac{p(\boldsymbol{\xi}, \boldsymbol{\eta} \vert \mathcal{D})}{p(\boldsymbol{\xi} \vert \boldsymbol{\eta},\mathcal{D})} \nonumber\\
&\propto&\frac{\mathcal{L}(\boldsymbol{\xi}; \mathcal{D}) p(\boldsymbol{\xi} \vert \boldsymbol{\eta}) p(\boldsymbol{\eta})}{ p(\boldsymbol{\xi} \vert \boldsymbol{\eta}, \mathcal{D})} \nonumber \\
&\propto& \frac{\exp\left(\ell(\boldsymbol{\xi}; \mathcal{D})\right) p(\boldsymbol{\xi} \vert \boldsymbol{\lambda}) \Bigg({\displaystyle \prod_{j=1}^q p(\lambda_j \vert \delta_j)} \Bigg) \Bigg({\displaystyle \prod_{j=1}^q p(\delta_j)} \Bigg)}{p(\boldsymbol{\xi} \vert \boldsymbol{\lambda}, \mathcal{D})}, \nonumber
\end{eqnarray}

\vspace{0.1cm}

\noindent where $\mathcal{L}(\boldsymbol{\xi}; \mathcal{D})$ is the likelihood function. An approximation $\tilde{p}(\boldsymbol{\eta} \vert \mathcal{D})$ to the above marginal posterior of $\boldsymbol{\eta}$ is obtained by substituting the Laplace approximation to $p(\boldsymbol{\xi} \vert \boldsymbol{\lambda}, \mathcal{D})$ (cf. Section 2.3) and by evaluating the resulting expression at the posterior mode $\hat{\boldsymbol{\xi}}_{\boldsymbol{\lambda}}$. Let us express the natural parameter in the generalized additive model as $\gamma_i=\varrho_i=\mathbf{b}^T_i \boldsymbol{\xi}$, with $\mathbf{b}^T_i$ the row vector corresponding to the $i${th} row of matrix $B$. Using the previous suggestion and noting that the determinant of the block diagonal matrix involved in the prior $p(\boldsymbol{\xi}\vert \boldsymbol{\lambda})$ is given by $\vert Q^{\boldsymbol{\lambda}}_{\boldsymbol{\xi}}\vert^{\frac{1}{2}}\propto \prod_{j=1}^q \lambda_j^{(K-1)/2}$, we obtain:

\vspace{-0.2cm}

\begin{eqnarray} \label{gam_tilde_eta}
\tilde{p}(\boldsymbol{\eta} \vert \mathcal{D})&\propto& \exp\Bigg(
\frac{1}{\varkappa} \sum_{i=1}^n \left[ y_i \mathbf{b}^T_i \hat{\boldsymbol{\xi}}_{\boldsymbol{\lambda}}-s\left(\mathbf{b}^T_i \hat{\boldsymbol{\xi}}_{\boldsymbol{\lambda}} \right)\right]-\frac{1}{2} \hat{\boldsymbol{\xi}}_{\boldsymbol{\lambda}}^T Q^{\boldsymbol{\lambda}}_{\boldsymbol{\xi}} \hat{\boldsymbol{\xi}}_{\boldsymbol{\lambda}} \Bigg) \nonumber \\
&&\times \ \Bigg(\prod_{j=1}^q \delta_j^{(\frac{\nu}{2}+a_{\delta}-1)} \exp\left(-\delta_j\Big(b_{\delta}+\frac{\nu}{2} \lambda_j\Big)\right)\Bigg)\ \Bigg(\prod_{j=1}^q \lambda_j^{\left( \frac{\nu+K-3}{2}\right)}\Bigg) \nonumber \\
&&\times \ \vert B^T \widetilde{W} B+Q^{\boldsymbol{\lambda}}_{\boldsymbol{\xi}} \vert^{-\frac{1}{2}}.
\end{eqnarray}

\noindent As Gamma priors have been chosen for the penalty parameters $\lambda_j$ and $\delta_j$, one recognizes in \eqref{gam_tilde_eta} the conditional conjugacy for $\delta_j$, as $\delta_j \vert \lambda_j, \mathcal{D} \sim \mathcal{G}\big(\frac{\nu}{2}+a_{\delta}, b_{\delta}+\frac{\nu}{2} \lambda_j\big)$. Under these prior specifications, the integration of \eqref{gam_tilde_eta} with respect to $\boldsymbol{\delta}$ is tractable and yields the (approximate) marginal penalty posterior:

\vspace{-0.2cm}

\begin{eqnarray} \label{gam_p_lambda}
\tilde{p}(\boldsymbol{\lambda} \vert \mathcal{D})&=&\int_{0}^{+\infty} \dots \int_{0}^{+\infty} \tilde{p}(\boldsymbol{\eta} \vert \mathcal{D}) \ d\delta_1\dots d\delta_q \nonumber \\
&\propto& \vert B^T \widetilde{W} B+Q^{\boldsymbol{\lambda}}_{\boldsymbol{\xi}} \vert^{-\frac{1}{2}} \exp\Bigg(
\frac{1}{\varkappa} \sum_{i=1}^n \left[ y_i \mathbf{b}^T_i \hat{\boldsymbol{\xi}}_{\boldsymbol{\lambda}}- s\left(\mathbf{b}^T_i \hat{\boldsymbol{\xi}}_{\boldsymbol{\lambda}} \right) \right]-\frac{1}{2} \hat{\boldsymbol{\xi}}_{\boldsymbol{\lambda}}^T Q^{\boldsymbol{\lambda}}_{\boldsymbol{\xi}} \hat{\boldsymbol{\xi}}_{\boldsymbol{\lambda}}\Bigg) \nonumber \\
&&\times \ \Bigg(\prod_{j=1}^q \lambda_j^{\left( \frac{\nu+K-3}{2}\right)}\Bigg) \Bigg( \prod_{j=1}^q \Big(b_{\delta}+\frac{\nu}{2} \lambda_j \Big)^{-(\frac{ \nu}{2}+a_{\delta})}\Bigg).
\end{eqnarray}

\vspace{0.4cm}

\noindent Applying a log transform on the penalty parameters $v_j=\log(\lambda_j),\  j=1,\dots,q$ and using the multivariate transformation method on (\ref{gam_p_lambda}), we obtain the following expression for the (log-) posterior of the log penalty vector:

\vspace{0.2cm}

\newcommand{\mtilde}{\widetilde{\mathcal{M}}_{\boldsymbol{\xi}}^{\mathbf{v}}}
\newcommand{\varpitilde}{\widetilde{\boldsymbol{\varpi}}}

\vspace{-0.4cm}

\begin{eqnarray} \label{gam_posterior_v}
\log \tilde{p}(\mathbf{v} \vert \mathcal{D}) &\dot{=}& -\frac{1}{2} \log \vert B^T \widetilde{W} B+Q_{\boldsymbol{\xi}}^{\mathbf{v}} \vert+\frac{\nu+K-1}{2}\sum_{j=1}^q v_j+\frac{1}{\varkappa} \sum_{i=1}^n y_i \mathbf{b}^T_i \hat{\boldsymbol{\xi}}_{\mathbf{v}} \nonumber \\
&&-\frac{1}{\varkappa} \sum_{i=1}^n s\left(\mathbf{b}^T_i \hat{\boldsymbol{\xi}}_{\mathbf{v}} \right)-\frac{1}{2} \hat{\boldsymbol{\xi}}_{\mathbf{v}}^T Q_{\boldsymbol{\xi}}^{\mathbf{v}} \hat{\boldsymbol{\xi}}_{\mathbf{v}}-\Big(\frac{ \nu}{2}+a_{\delta}\Big)\sum_{j=1}^q \log\Big(b_{\delta}+\frac{\nu}{2} \exp(v_j) \Big),
\end{eqnarray}

\noindent where $Q_{\boldsymbol{\xi}}^{\mathbf{v}}$ is the symmetric block diagonal matrix:

\vspace{-0.3cm}

\begin{eqnarray}
Q_{\boldsymbol{\xi}}^{\mathbf{v}}=\begin{pmatrix}
\zeta I_{p+1} & 0_{p+1,q\times(K-1)} \\
0_{q\times(K-1),p+1} & \text{diag}(\exp(v_1),\dots,\exp(v_q)) \otimes P
\end{pmatrix} \nonumber 
\end{eqnarray}

\vspace{0.2cm}

\noindent and $\hat{\boldsymbol{\xi}}_{\mathbf{v}}:=\left(B^T \widetilde{W}B+Q_{\boldsymbol{\xi}}^{\mathbf{v}}\right)^{-1} \widetilde{\boldsymbol{\varpi}}$. The gradient $\nabla_{\mathbf{v}} \log \tilde{p}(\mathbf{v} \vert \mathcal{D})$ and Hessian $\nabla_{\mathbf{v}}^2 \log \tilde{p}(\mathbf{v} \vert \mathcal{D})$ of expression (\ref{gam_posterior_v}) can be analytically derived, see Appendix (A1) for full details. These expressions will turn to be useful to explore the marginal posterior of the penalty parameters.

\subsection{Strategy to explore the posterior penalty space}

\noindent An approximation to the marginal posterior of the latent variables $\boldsymbol{\xi}$ (including the regression and spline parameters in the generalized additive model) can be obtained by integrating out the penalty parameters as in (\ref{integrate_out}). Obtaining such a quadrature requires to explore the posterior of the penalty parameters $\boldsymbol{\lambda}=\exp(\mathbf{v})$.\\
\indent Two strategies are suggested according to the dimension $q$ of the penalty vector. When $q$ is small or moderate (say $q\leq 4$), a grid strategy is proposed that is sensitive to asymmetries in the response surface $\tilde{p}(\mathbf{v} \vert \mathcal{D})$, with the skew-normal family of distributions forming the backbone to handle asymmetry. As the computational cost of constructing a grid grows with dimension $q$, we suggest an alternative strategy relying on MCMC to draw a set of points in the domain of the posterior of the penalty parameters when $q$ is large.\\
\noindent This hybrid approach alternates between a deterministic grid and a sampling scheme, giving to the end-user a complete and rapid tool to fit GAMs in a full Bayesian framework even when the number of smooth functions is large. A preliminary milestone for both strategies is to find the posterior mode $\hat{\mathbf{v}}$ of $\log \tilde{p}(\mathbf{v} \vert \mathcal{D})$ as it represents the ``center of gravity'' around which the exploration will depart. To this end, a Newton-Raphson algorithm is implemented in which we take advantage of the analytical forms for the gradient and Hessian of $\log \tilde{p}(\mathbf{v} \vert \mathcal{D})$ to speed up the computational process. Once $\hat{\mathbf{v}}$ is obtained, we proceed with posterior exploration.

\subsubsection{Grid-strategy with skew-normal match when $q$ is small}

\noindent An elementary approach to explore $\tilde{p}(\mathbf{v} \vert \mathcal{D})$ could rely on a multivariate Gaussian approximation to the posterior of the log penalty parameters $\mathbf{v}$, i.e. $\tilde{p}_G(\mathbf{v}\vert \mathcal{D})= \mathcal{N}_{\text{dim}({\mathbf{v}})}\Big(\hat{\mathbf{v}}, \big(-\mathcal{H}^*\big)^{-1}\Big)$, where the covariance matrix is obtained from the Hessian $\mathcal{H}^*=\nabla_{\mathbf{v}}^2 \log \tilde{p}(\hat{\mathbf{v}} \vert \mathcal{D})$ evaluated at the mode $\hat{\mathbf{v}}$. However, as already pointed in \cite{martins2013bayesian}, the presence of potential asymmetries would not be captured by a Gaussian approximation. Instead, to efficiently explore the posterior penalty space, a grid strategy is proposed, which implicitly takes into account asymmetries by using skew-normal distributions to approximate the conditional posterior of each penalty parameter through a moment-matching approach.\\
\indent  The skew-normal family was first introduced by \cite{azzalini1985class}, see \cite{azzalini2013skew} for more details. In the univariate case, a random variable $X$ has a skew-normal distribution denoted by $X\sim \text{SN}(\mu, \varsigma^2, \rho)$ if its probability density function at $x \in \mathbb{R}$ is:

\vspace{-0.2cm}

\begin{eqnarray} \label{skew_normal_azzalini}
p(x)=\frac{2}{\varsigma} \varphi \left(\frac{x-\mu}{\varsigma} \right)\ \Phi\left(\rho \frac{(x-\mu)}{\varsigma}\right),
\end{eqnarray}

\begin{figure}[h!]
\centering
\includegraphics[width=16.5cm, height=7.5cm]{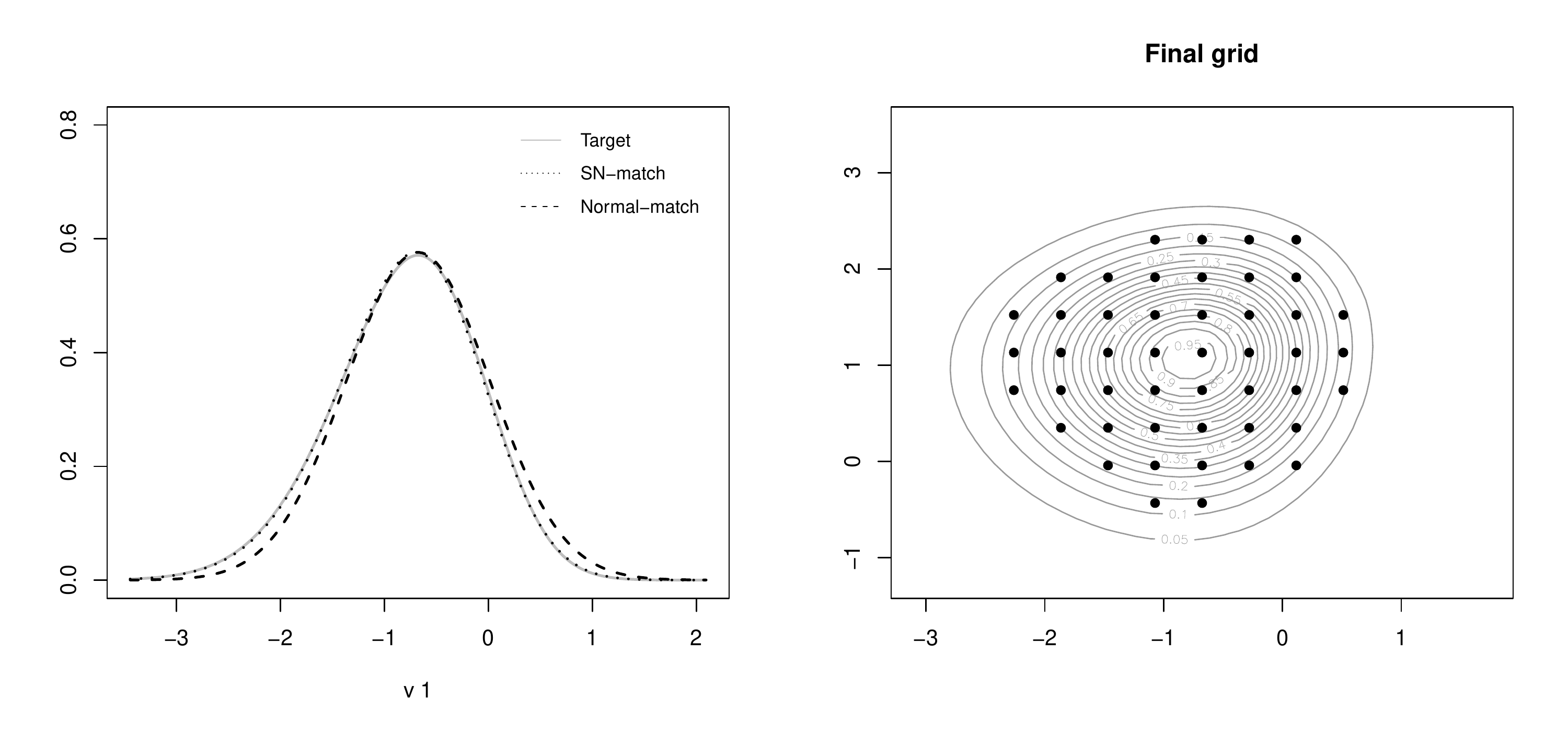}
\caption{Left: Skew-normal fit (dotted) and naive Gaussian match (dashed) to the conditional $\tilde{p}(v_1 \vert \hat{v}_{2},\mathcal{D})$ (gray). The skew-normal fit is closer to the target and captures the lack of symmetry present in the target. Right: Final grid construction to explore $\tilde{p}(\mathbf{v} \vert \mathcal{D})$.}
\label{fig:figure1} 
\end{figure}

\vspace{0.3cm}

\noindent where $\mu \in \mathbb{R}$ is a location parameter, $\varsigma \in \mathbb{R}_{+}$ a scale parameter and $\rho\in \mathbb{R}$ a shape parameter regulating skewness. Also, $\varphi(\cdot)$ and $\Phi(\cdot)$  denote the standard Gaussian density function and its cumulative distribution function respectively, such that setting $\rho=0$ yields the $\mathcal{N}(\mu, \varsigma^2)$ distribution. We suggest to approximate the conditional posterior distribution of $(v_j |\hat{\mathbf{v}}_{-j},\mathcal{D})$ $(j=1,...,q)$ with a skew-normal distribution by matching its first three empirical moments with the theoretical ones for the density in (\ref{skew_normal_azzalini}), where $\hat{\mathbf{v}}_{-j}$ denotes the vector $\hat{\mathbf{v}}$ without the $j$th entry. Appendix (A2) shows the derivations to  obtain $\mu^*, \varsigma^*$ and $\rho^*$ in the approximating skew-normal distribution $\text{SN}_j(\mu^*, \varsigma^{*2},\rho^*)$ through moment matching. \\
\indent Once a skew-normal distribution $\text{SN}_j(\mu^*, \varsigma^{*2},\rho^*)$ has been adjusted to the conditional $\tilde{p}(v_j \vert \hat{\mathbf{v}}_{-j},\mathcal{D})$, we construct an equidistant grid $\{v_{jm}\}_{m=1}^M$ of size $M$ from the $2.5${th} to the $97.5${th} quantiles of the skew-normal fit denoted by $\text{SN}_{j,0.025}$ and $\text{SN}_{j,0.975}$ respectively. This process is repeated across all dimensions $j=1,\dots,q$ and a Cartesian product of the univariate grids is taken, ending up with a total of $M^q$ (multivariate) grid points. Next, a filtering strategy is implemented to get rid of quadrature points associated to a small posterior mass. Let us consider the normalized posterior $R(\mathbf{v})=\tilde{p}(\mathbf{v} \vert \mathcal{D})/\tilde{p}(\hat{\mathbf{v}} \vert \mathcal{D})$ and use the property that $-2\log R(\mathbf{v})$ is approximately distributed as a chi-square distribution with $\text{dim}(\mathbf{v})$ degrees of freedom denoted by $\chi^2_{\text{dim}(\mathbf{v})}$. Then, an approximate $(1-\alpha)$ credible region for $\mathbf{v}$ is defined by the set of values in $\mathbb{R}^{\text{dim}(\mathbf{v})}$ such that $R(\mathbf{v})\geq \exp\left( -.5 \chi^2_{\text{dim}(\mathbf{v});1-\alpha}\right)$. As an illustration, take $\alpha=0.05$ and $\text{dim}(\mathbf{v})=2$. If we decide to concentrate on quadrature points in the 95$\%$ credible region for $\mathbf{v}$, then the preceding result would suggest to discard values $\mathbf{v}$ in the bivariate grid for which $R(\mathbf{v})<\exp(-.5 \chi^2_{2;0.95})=.05$, leaving $\widetilde{M}$ grid points. Figure \ref{fig:figure1} highlights the skew-normal match and the final grid in an example with $q=2$ nonlinear smooth functions in the additive predictor and data generated from a Poisson response with sample size $n=250$.


\vspace{-0.2cm}

\subsubsection{Independence sampling when $q$ is large}

\noindent When the number of smooth functions $q$ in the additive model is above a certain threshold (say $q>4$), the preceding computational strategy becomes too demanding as the number of quadrature points (following from the Cartesian product of the grid points for each penalty parameter $\exp(v_j)$ $(j=1,\ldots,q)$) explodes. A cost-effective alternative relies on MCMC to sample values from the posterior $\tilde{p}(\mathbf{v} \vert \mathcal{D})$. More thoroughly, an independence sampler is implemented using a multivariate Student-$t$ proposal distribution $ t_{\vartheta}(\hat{\mathbf{v}},(-\mathcal{H}^{*})^{-1})$ with density $h(\mathbf{v}\vert \hat{\mathbf{v}})$, degrees of freedom  ($\vartheta=3$, say), a mean set at the posterior mode $\hat{\mathbf{v}}$, and variance-covariance matrix $(\vartheta/(\vartheta-2))(-\mathcal{H}^{*})^{-1}$, see Section 2.5.1.\\
\indent Algorithm 1 summarizes the strategy to explore $\tilde{p}(\mathbf{v} \vert \mathcal{D})$. When $q \leq 4$, a grid is constructed using a Cartesian product of marginal grids delimited by quantiles of approximating skew-normal densities. Exploration in larger dimensions relies on the independence Metropolis-Hastings sampler. This algorithm will be used in the next section to approximate the marginal posterior of the latent field.

\newpage 

\hrule
\vspace{0.1cm}
\noindent {\textbf{Algorithm 1: Exploration of}\ $\tilde{p}(\mathbf{v} \vert \mathcal{D})$} \label{grid_algorithm_skew_normal}
\vspace{0.1cm}
\hrule 
\vspace{0.2cm}
\noindent 1:\ \textbf{If} $q\leq 4$\ \textbf{do}\ \text{(Grid strategy, cf. Section 2.5.1)}\\
\noindent 2:\hspace{0.4cm} \textbf{for}\ $j=1,\dots,q$\ \textbf{do}\\
3:\hspace{1cm} \text{Compute the skew-normal match}\ $\text{SN}_j(\mu^*, \varsigma^{*2},\rho^*)$ \ \text{to}\ $\tilde{p}(v_j \vert \hat{\mathbf{v}}_{-j}, \mathcal{D})$.\\
4:\hspace{1cm} \text{Construct a Cartesian grid}\ $\{v_{jm}\}_{m=1}^M$\ \text{from}\ $\text{SN}_{j,0.025}$\  \text{to}\ $\text{SN}_{j,0.975}$. \\ 
5:\hspace{0.4cm} \textbf{end for} \\
6:\hspace{0.4cm} \text{Compute the Cartesian product of the univariate grids}\  $\mathcal{C}=\times_{j=1}^q \{v_{jm}\}_{m=1}^M$. \\
7:\hspace{0.4cm} \text{Choose}\  $\alpha$\ \text{and keep the} $\widetilde{M}$ \ \text{values in }\  $\mathcal{C}$\  \text{such that}\ $R(\mathbf{v})\geq \exp\left(-.5 \chi^2_{q;1-\alpha}\right)$.\\
8:\ \textbf{else do}\ \text{(Independence sampling, cf. Section 2.5.2)}\\
9:\hspace{0.4cm} \text{Choose an initial value}\  $\mathbf{v}^{(0)}=\hat{\mathbf{v}}$.\\
10:\hspace{0.25cm} \textbf{for}\  $m=1,\dots,\widetilde{M}$\ \textbf{do}\\
11:\hspace{1cm} \text{Generate}\  $\mathbf{v}^{(\text{prop})} \sim h(\mathbf{v}\vert \hat{\mathbf{v}})$.\\
12:\hspace{1cm} \text{Compute the acceptance probability}\  $\alpha=\min\left(1, \frac{\tilde{p}\big(\mathbf{v}^{(\text{prop})} \vert \mathcal{D}\big)  h\big(\mathbf{v}^{(m-1)}\vert \hat{\mathbf{v}}\big)}{ \tilde{p}\big(\mathbf{v}^{(m-1)} \vert \mathcal{D}\big) h\big(\mathbf{v}^{(\text{prop})}\vert \hat{\mathbf{v}}\big)}\right)$.\\
13:\hspace{1cm} \text{Draw}\ $u\sim\mathcal{U}(0,1)$.\\
14:\hspace{1cm} \text{If} $u\leq\alpha$, set $\mathbf{v}^{(m)}=\mathbf{v}^{(\text{prop})}$, else set $\mathbf{v}^{(m)}=\mathbf{v}^{(m-1)}$.\\
15:\hspace{0.25cm} \textbf{end for}
\vspace{0.2cm}
\hrule

\vspace{0.4cm}

\subsection{Posterior inference on the latent field}

\subsubsection{Approximate latent field posterior}

\noindent Using the Laplace approximation discussed in Section 2.3, the posterior of the latent vector $\boldsymbol{\xi}$ can be obtained as follows:

\vspace{-0.6cm}

\begin{eqnarray} \label{posterior_latent_gam}
p(\boldsymbol{\xi}\vert \mathcal{D})&=& \int_{\mathbb{R}^{q}_{++}}  p(\boldsymbol{\xi} \vert \boldsymbol{\lambda},  \mathcal{D})\ p(\boldsymbol{\lambda} \vert \mathcal{D})\ d\boldsymbol{\lambda} \nonumber \\
&\approx& \int_{\mathbb{R}^{q}_{++}}  \tilde{p}_G(\boldsymbol{\xi} \vert \boldsymbol{\lambda},  \mathcal{D})\ \tilde{p}(\boldsymbol{\lambda} \vert \mathcal{D})\ d\boldsymbol{\lambda} \nonumber \\
&\approx& \int_{\mathbb{R}^{q}} \tilde{p}_G(\boldsymbol{\xi} \vert \exp(\mathbf{v}),  \mathcal{D})\ \tilde{p}(\mathbf{v} \vert \mathcal{D})\ d\mathbf{v},
\end{eqnarray}

\noindent where the last line follows from the change of variable in log-scale. Using Algorithm 1, we get a set of quadrature points $\{\mathbf{v}^{(m)}\}_{m=1}^{\widetilde{M}}$. Defining:

\vspace{-0.3cm}

\begin{eqnarray} \label{gam_weights}
\omega_m= \frac{\tilde{p}(\mathbf{v}^{(m)} \vert \mathcal{D})}{\sum_{m=1}^{\widetilde{M}}\tilde{p}(\mathbf{v}^{(m)} \vert \mathcal{D})},\ \ m=1,\dots,\widetilde{M},
\end{eqnarray}

\vspace{0.2cm}

\noindent when $q\leq4$ and $\omega_m=1/{\widetilde{M}}$ otherwise, Equation \eqref{posterior_latent_gam} suggests to approximate $p(\boldsymbol{\xi} |\mathcal{D})$ by:

\vspace{-0.6cm}

\begin{eqnarray} \label{approx_post_latent_gam}
\tilde{p}(\boldsymbol{\xi} \vert \mathcal{D})=\sum_{m=1}^{\widetilde{M}}\ \omega_m\  \mathcal{N}_{\text{dim}(\boldsymbol{\xi})}\left(\hat{\boldsymbol{\xi}}_{\mathbf{v}^{(m)}}, \widehat{\Sigma}_{\mathbf{v}^{(m)}}\right), 
\end{eqnarray}

\noindent where $\hat{\boldsymbol{\xi}}_{\mathbf{v}^{(m)}}=\left(B^T \widetilde{W}B+Q_{\boldsymbol{\xi}}^{\mathbf{v}^{(m)}} \right)^{-1} \widetilde{\boldsymbol{\varpi}}$ and $\widehat{\Sigma}_{\mathbf{v}^{(m)}}=\left(B^T \widetilde{W}B+Q_{\boldsymbol{\xi}}^{\mathbf{v}^{(m)}} \right)^{-1}$ are the conditional posterior mode and variance-covariance matrix resulting from the iterative Laplace approximations proposed in Section 2.3. Note that the computational cost of reevaluating the conditional posterior mode and variance-covariance for each penalty $\exp(\mathbf{v}^{(m)})$ in the grid can be reduced by adding an extra layer of approximation by replacing $\widetilde{W}$ in the Newton-Raphson procedure by its value $\widetilde{W}_{\hat{\mathbf{v}}}$ at the posterior mode. A point estimate for the latent vector is given by the posterior mean of (\ref{approx_post_latent_gam}), which is a mixture of the location components, i.e. $
\hat{\boldsymbol{\xi}}=\sum_{m=1}^{\widetilde{M}} \omega_m\  \hat{\boldsymbol{\xi}}_{\mathbf{v}^{(m)}}$.

\vspace{-0.1cm}

\subsubsection{Credible intervals for latent elements and additive terms}

Approximate pointwise credible intervals for latent elements $\xi_h,\ h=1,\dots,\text{dim}({\boldsymbol{\xi}})$ can be straightforwardly obtained by starting from the finite mixture given in (\ref{approx_post_latent_gam}). The approximate posterior for the $h$th latent element is $\tilde{p}(\xi_h\vert \mathcal{D})= \sum_{m=1}^{\widetilde{M}} \omega_m\ \mathcal{N}_1\Big(\hat{\xi}_{h,\mathbf{v}^{(m)}},\widehat{\Sigma}_{hh,\mathbf{v}^{(m)}}\Big)$, where $\hat{\xi}_{h,\mathbf{v}^{(m)}}$ is the $h$th entry of vector $\hat{\boldsymbol{\xi}}_{\mathbf{v}^{(m)}}$ and $\widehat{\Sigma}_{hh,\mathbf{v}^{(m)}}$ is the $h$th entry on the diagonal of matrix $\widehat{\Sigma}_{\mathbf{v}^{(m)}}$. The latter expression can be used to construct a $(1-\alpha) \times 100\%$ quantile-based credible interval for $\xi_h$. To obtain pointwise set estimates of a smooth function $f_j$, let $\{x_{l}\}_{l=1}^L$ be an equidistant (fine) grid on the domain of $f_j$ and $\boldsymbol{\xi}_{\boldsymbol{\theta}_j}$ be the subvector of the latent field corresponding to the spline vector $\boldsymbol{\theta_j}=(\theta_{j1},\dots,\theta_{jK-1})^T$. Also, denote by $\widetilde{\mathbf{b}}_l$ the vector of B-splines in the basis evaluated at $x_l$. The function $f_j$ at point $x_{l}$ is thus modeled as  $f_j(x_l \vert \boldsymbol{\xi}_{\boldsymbol{\theta}_j})=\widetilde{\mathbf{b}}_l^T \boldsymbol{\xi}_{\boldsymbol{\theta}_j}$ and from (\ref{approx_post_latent_gam}) the posterior of $\boldsymbol{\xi}_{\boldsymbol{\theta}_j}$ is approximated by the finite mixture:

\vspace{-0.8cm}

\begin{eqnarray} \label{post_mix_theta_j}
\tilde{p}(\boldsymbol{\xi}_{\boldsymbol{\theta}_j} \vert \mathcal{D})=\sum_{m=1}^{\widetilde{M}} \omega_m\  \mathcal{N}_{K-1}\left(\hat{\boldsymbol{\xi}}_{\boldsymbol{\theta}_{j},\mathbf{v}^{(m)}},\widehat{\Sigma}_{\boldsymbol{\theta}_{j},\mathbf{v}^{(m)}}\right),
\end{eqnarray}

\vspace{0.1cm}

\noindent where $\widehat{\Sigma}_{\boldsymbol{\theta}_{j},\mathbf{v}^{(m)}}$ is a submatrix of $\widehat{\Sigma}_{\mathbf{v}^{(m)}}$ corresponding to the variance-covariance matrix of $\boldsymbol{\xi}_{\boldsymbol{\theta}_j}$. As $f_j(x_l \vert \boldsymbol{\xi}_{\boldsymbol{\theta}_j})$ is a linear combination of the spline vector, a natural candidate to approximate the posterior $p(f_j(x_l \vert \boldsymbol{\xi}_{\boldsymbol{\theta}_j}) \vert \mathcal{D})$ is to use a mixture of univariate normals:

\vspace{-0.3cm}

\begin{eqnarray}
\tilde{p}(f_j(x_l \vert \boldsymbol{\xi}_{\boldsymbol{\theta}_j}) \vert \mathcal{D})=\sum_{m=1}^{\widetilde{M}} \omega_m\  \mathcal{N}_1\Big(\widetilde{\mathbf{b}}_l^T \hat{\boldsymbol{\xi}}_{\boldsymbol{\theta}_{j},\mathbf{v}^{(m)}},\widetilde{\mathbf{b}}_l^T \widehat{\Sigma}_{\boldsymbol{\theta}_{j},\mathbf{v}^{(m)}} \widetilde{\mathbf{b}}_l\Big). \nonumber
\end{eqnarray}

\vspace{0.2cm}

\noindent A quantile-based credible interval for $f_j$ at point $x_l$ can easily be computed from the above (approximate) univariate posterior.

\vspace{-0.2cm}

\section{Simulations}

The performance of the LPS approach (with cubic B-splines and a third order penalty) is assessed through different simulation scenarios and compared with results obtained using the {\tt{gam}} function from the {\ttfamily{mgcv}} package in {\bf{R}} (\citealp{wood2017generalized}), a popular and established toolkit for estimating GAMs. Options of the {\tt{gam}} function are carefully chosen so that the generated results can be meaningfully compared to these obtained using our Laplace-P-spline approach. In particular, smooth terms are specified with the {\tt{gam}} function using {\textit{s(x, bs=``ps", k=K, m=c(2,3))} , where \textit{x} is the vector of covariate values associated to the estimated smooth function and {\textit{ps}} specifies a P-spline basis. The scalar \textit{k} is the basis dimension, the first entry in $\textit{m}=c(\cdot,\cdot)$ refers to the order of the spline basis (with order 2 corresponding to cubic P-splines), while the second entry refers to the order of the difference penalty. Another chosen option in {\tt{gam}} is $method=``REML"$, requiring an estimation of the penalty parameters $\boldsymbol{\lambda}$ by restricted maximum likelihood. It corresponds to an empirical Bayes approach in the sense that a Bayesian log marginal likelihood is maximized with respect to $\boldsymbol{\lambda}$ in a context where penalties come from Gaussian priors on the spline coefficients (\citealp{marra2011practical}; \citealp{wood2013straightforward}).  The optimization method in {\tt{gam}} is chosen to be \textit{optimizer=c(``outer",``newton")} as it provides reliable and stable computations.
		
\vspace{-0.2cm}		
		
\subsection{Estimation of the regression parameters in the linear part}
\label{sec:6.1}

\noindent The simulation setting entails $S=500$ replications of a data set of size $n=300$ with three covariates in the linear part generated independently as  $z_{i1}\sim \text{Bern}(0.5)$, $z_{i2}\sim \mathcal{N}(0,1)$ and $z_{i3}\sim \mathcal{N}(0,1)$, for $i=1,\dots,n$ and coefficients $\beta_0=-1.50,\  \beta_1=0.70,\ \beta_2=-0.80,\ \beta_3=0.40$. The covariates for the smooth functions are independent draws from the Uniform distribution on the domain $[-1,1]$. The smooth additive terms coincide with the functions:

\vspace{-0.5cm}

\begin{eqnarray}
f_1(x_1)&=&-4x_1^6+2x_1^2+\cos(2\pi x_1)-0.1, \nonumber \\
f_2(x_2)&=&3x_2^5+2 \sin(4x_2)+1.5x_2^2-0.5, \nonumber \\
f_3(x_3)&=&\sin(3\pi x_3). \nonumber
\end{eqnarray}

\vspace{0cm}

\noindent The above functions are specified as a linear combination of cubic B-splines with a third order penalty and $K=15$ B-splines in $[-1,1]$.  The frequentist properties of the Bayesian estimators are measured by the bias, the empirical standard error (ESE), the root mean square error (RMSE) and coverage probability (CP) of the $90\%$ and $95\%$ (pointwise) credible intervals for the linear coefficients. Four scenarios are considered for the response variable, namely (I) Generation from a Poisson distribution $y_i \sim \text{Poisson}(\mu_i)$, with $\mu_i=\exp(\varrho_i)$ to illustrate the case of count data, (II) Generation from a Gaussian $y_i \sim \mathcal{N}(\mu_i, \sigma^2=0.3)$, with $\mu_i=\varrho_i$, (III) Generation from a Binomial $y_i \sim \text{Bin}(15, p_i)$ and (IV) Generation from a Bernoulli $y_i\sim \text{Bern}(p_i)$ to illustrate the case of binary responses with success probability $p_i=\exp(\varrho_i)/(1+\exp(\varrho_i))$ for the Binomial and Bernoulli cases.\\
\indent Table \ref{table:tab1} shows the simulation results and comparisons with the {\tt{gam}} function. For all the considered data types, the Laplace-P-spline approach exhibits nonsignificant biases and the estimated coverage probabilities are consistent with their nominal level. Also, the ESE and RMSE show a behavior comparable to what is observed with the {\tt{gam}} output. For the Bernoulli scenario, ESEs are smaller with LPS, but biases are slightly larger than with {\tt{gam}}. The frequentist coverage of credible intervals remain compatible whatever the method used. A notable feature of the Laplace-P-spline methodology is that it requires a low computational cost despite being fully Bayesian. In fact, our algorithm (underlying a fully Bayesian approach) is purely written in {\bf{R}} (without any parallelization) and takes approximately 0.9 seconds per dataset in the above scenario as compared to 0.05 seconds for the {\tt{gam}} function (coding an empirical Bayes approach) for simulations performed on a machine equipped with an Intel Xeon E-2186M CPU running at a clock speed of 2.90 GHz. Considering that the {\tt{gam}} algorithm is neither fully Bayesian nor entirely written in {\bf{R}} (as most of the script relies on {\bf{C}} code which is much faster), the Laplace-P-spline toolkit can be considered a serious competitor for approximate full Bayesian inference in GAMs when smooth functions are modeled with P-splines.

\newpage 


\begin{center}
\begin{table}[h!]
\linespread{0.7} \selectfont
\setlength{\tabcolsep}{2.7pt}
\begin{tabular}{lcccccc}
\hline \\
Data &  Parameters &  Bias  &  CP$_{90\%}$ &  CP$_{95\%}$ &  ESE & RMSE \\
\hline \\
\vspace{0.1cm}
\phantom{k} & $\beta_1=\phantom{-}0.70$ & \hspace{0.2cm} \phantom{-}0.001 (\phantom{-}0.003) & \hspace{0.1cm} 87.4 (88.2) & \hspace{0.1cm} 94.0 (94.6) & 0.122 (0.122) & 0.122 (0.121) \\
\vspace{0.1cm}
Poisson & $\beta_2=-0.80$ & \hspace{0.2cm} \phantom{-}0.006 (\phantom{-}0.003) & \hspace{0.1cm} 91.0 (90.8) & \hspace{0.1cm} 95.8 (95.6) & 0.061 (0.061) & 0.062 (0.061) \\
\vspace{0.1cm}
\phantom{k} & $\beta_3=\phantom{-}0.40$ & \hspace{0.2cm} -0.001 (\phantom{-}0.000) & \hspace{0.1cm} 90.0 (90.0) & \hspace{0.1cm} 95.8 (96.4) & 0.060 (0.060) & 0.060 (0.059) \\
\hline \\
\vspace{0.1cm}
\phantom{k} & $\beta_1=\phantom{-}0.70$ & \hspace{0.2cm} \phantom{-}0.001 (\phantom{-}0.001) & \hspace{0.1cm} 90.6 (90.0) & \hspace{0.1cm} 96.4 (96.4) & 0.065 (0.065) & 0.065 (0.065) \\
\vspace{0.1cm}
Normal  & $\beta_2=-0.80$ & \hspace{0.2cm} -0.001 (-0.001) & \hspace{0.1cm} 89.0 (89.4) & \hspace{0.1cm} 94.8 (95.0) & 0.033 (0.033) & 0.033 (0.033) \\
\vspace{0.1cm}
\phantom{k} & $\beta_3=\phantom{-}0.40$ & \hspace{0.2cm} \phantom{-}0.000 (\phantom{-}0.000) & \hspace{0.1cm} 89.6 (90.2) & \hspace{0.1cm} 94.8 (95.2) & 0.034 (0.034) & 0.033 (0.034) \\
\hline \\
\vspace{0.1cm}
\phantom{k} & $\beta_1=\phantom{-}0.70$ & \hspace{0.2cm} \phantom{-}0.004 (\phantom{-}0.006) & \hspace{0.1cm} 89.8 (90.8) & \hspace{0.1cm} 94.8 (95.0) & 0.090 (0.090) & 0.090 (0.091) \\
\vspace{0.1cm}
Binomial & $\beta_2=-0.80$ & \hspace{0.2cm} \phantom{-}0.011 (\phantom{-}0.008) & \hspace{0.1cm} 88.8 (88.6) & \hspace{0.1cm} 93.6 (94.2) & 0.047 (0.048) & 0.049 (0.048) \\
\vspace{0.1cm}
\phantom{k} & $\beta_3=\phantom{-}0.40$ & \hspace{0.2cm} -0.003 (-0.001) & \hspace{0.1cm} 92.6 (92.6) & \hspace{0.1cm} 96.4 (96.8) & 0.042 (0.042) & 0.042 (0.042) \\
\hline \\
\vspace{0.1cm}
\phantom{k} & $\beta_1=\phantom{-}0.70$ & \hspace{0.2cm} -0.077 (-0.008) & \hspace{0.1cm} 87.4 (87.8) & \hspace{0.1cm} 93.0 (93.0) & 0.320 (0.349) & 0.329 (0.349) \\
\vspace{0.1cm}
Bernoulli & $\beta_2=-0.80$ & \hspace{0.3cm} 0.082 (\phantom{-}0.005) & \hspace{0.1cm} 87.6 (91.8) & \hspace{0.1cm} 93.0 (96.4) & 0.155 (0.175) & 0.175 (0.174) \\
\vspace{0.1cm}
\phantom{k} & $\beta_3=\phantom{-}0.40$ & \hspace{0.2cm} -0.038 (\phantom{-}0.003) & \hspace{0.1cm} 88.6 (89.8) & \hspace{0.1cm} 93.2 (94.0) & 0.159 (0.176) & 0.163 (0.176) \\
\hline\\
\end{tabular}
\caption{Simulation results with the LPS method for $S=500$ replicates of sample size $n=300$ for different types of response (Poisson, Normal, Binomial and Bernoulli). The values in parentheses are estimation results from the {\tt{gam}} function.}
\label{table:tab1}
\end{table}
\end{center}

\vspace{-1.5cm}

\subsection{Estimation of the additive terms $f_j$}

\noindent The coverage properties of approximate $90\%$ pointwise credible intervals for the additive terms $f_1, f_2$ and $f_3$ are reported in Table \ref{table:tab2} for selected values of the covariate on $[-1,1]$. An asterisk superscript is added to the estimated coverage to indicate incompatibility with the nominal value. Results of the {\tt{gam}} function are labeled ``MGCV''. In addition to the LPS approach, Table \ref{table:tab2} also highlights the coverage performance of LPSMAP, where each penalty parameter is replaced by its posterior mode $\hat{\boldsymbol{\lambda}}=\exp(\hat{\mathbf{v}})$ in our Laplace-P-spline method. For LPSMAP the uncertainty in the selection of $\boldsymbol{\lambda}$ is ignored (as in Wood's approach), such that the mixture in Equation \eqref{post_mix_theta_j} is omitted and the point estimate of the latent vector and its associated variance-covariance matrix become $\hat{\boldsymbol{\xi}}_{\hat{\mathbf{v}}}=\left(B^T \widetilde{W}B+Q_{\boldsymbol{\xi}}^{\hat{\mathbf{v}}} \right)^{-1} \widetilde{\boldsymbol{\varpi}}$ and $\widehat{\Sigma}_{\hat{\mathbf{v}}}=\left(B^T \widetilde{W}B+Q_{\boldsymbol{\xi}}^{\hat{\mathbf{v}}} \right)^{-1}$ respectively. With LPSMAP, an approximate $(1-\alpha) \times 100\%$ credible interval for function $f_j$ at point $x_l$ is computed from a frequentist perspective, $\hat{f}_j(x_l)\pm z_{\alpha/2} \sqrt{\widetilde{\mathbf{b}}_l^T\widehat{\Sigma}_{\boldsymbol{\theta}_{j},\hat{\mathbf{v}}} \widetilde{\mathbf{b}}_l}$.\\
\indent As can be seen from Table \ref{table:tab2}, the LPS and LPSMAP methods perform well in the Poisson, Normal and Binomial scenarios as estimated frequentist coverage probabilities are close to the nominal level at almost all selected covariate values. The {\tt{gam}} results also show a similar performance across all scenarios. Comparing LPS and LPSMAP, we observe that omitting the penalty uncertainty globally translates into a slight decrease in percentage points for the estimated coverage probability. Yet, the LPSMAP approach still exhibits close to nominal coverage for all the functions. In terms of computational speed, the LPSMAP approach is approximately four times faster than the LPS approach and four times slower than {\tt{gam}} ($\approx$ 0.07 seconds vs 0.28 seconds).

\newpage 


\phantom{k}
\vspace{1.6cm}

\begin{table}[h!]
\linespread{0.7} \selectfont
\setlength{\tabcolsep}{2.6pt}
\begin{tabular}{lclccccccccc}
\hline \\
Data &  $f$ &  Method &  -0.95  &  -0.70 &  -0.50 &   -0.20 &  \hspace{0.1cm} 0.00 &   0.20 &   0.50 &   0.70 &  0.95\\
\hline
\\
\vspace{0.1cm}
\vphantom & \hspace{0.05cm}  $f_1$ & LPS &  \hspace{0.05cm} 86.0$^{*}$ &  89.8   &  91.6  &   91.2 &  88.2  &  91.4 &  87.0  &  88.4 &  87.6 \\
\vspace{0.1cm}
\vphantom & \hspace{0.05cm}  $f_1$ & LPSMAP & \hspace{0.05cm} 85.8$^{*}$ &  89.2   &  89.6 &  90.8 & 88.2 &  91.4 & \hspace{0.05cm} 86.0$^{*}$  &  87.6 &  87.0 \\
\vspace{0.1cm}
\vphantom & \hspace{0.05cm}  $f_1$ & MGCV &  87.8 & 91.6 &  92.0  & 90.6 &  90.6  &  92.0 &  89.4  &  92.2 &  89.0 \\
\vspace{0.1cm}
\vphantom{} Poisson & \hspace{0.05cm}  $f_2$ & LPS &  93.2 & \hspace{0.05cm} 82.8$^{*}$   &     89.2  &  \hspace{0.05cm} 84.4$^{*}$ &  91.2 &  89.2 & \hspace{0.05cm} 86.2$^{*}$ &  92.6 &  87.4  \\
\vspace{0.1cm}
\vphantom & \hspace{0.05cm}  $f_2$ & LPSMAP &  \hspace{-0.06cm} 92.4 &  \hspace{0.05cm} 81.4$^{*}$   &  87.4  &  \hspace{0.06cm} 81.4$^{*}$ &  90.2  &  89.0 & \hspace{0.05cm} 85.2$^{*}$  &  92.4 &  86.8 \\
\vspace{0.1cm}
\vphantom & \hspace{0.05cm}  $f_2$ & MGCV &   92.6 &  87.6   &  90.8  &   89.8 &  92.4  & 91.0 & 89.8  & 92.2 & 89.0 \\
\vspace{0.1cm}
\vphantom & \hspace{0.05cm}  $f_3$ & LPS &   88.8 & 87.2 & \hspace{0.05cm}  86.0$^{*}$ &  87.6 & 90.2 & \hspace{0.05cm}  86.0$^{*}$ &  \hspace{0.05cm}  86.0$^{*}$ &  89.2 &  90.6 \\
\vspace{0.1cm}
\vphantom & \hspace{0.05cm}  $f_3$ & LPSMAP &  \hspace{-0.08cm} 88.8 &  87.2   &  \hspace{0.06cm} 86.2$^{*}$  &   87.6 &  90.2  & \hspace{0.06cm} 85.8$^{*}$ & \hspace{0.05cm} 86.0$^{*}$  &  89.4 &  90.6 \\
\vspace{0.1cm}
\vphantom & \hspace{0.05cm}  $f_3$ & MGCV &  90.4 &  88.6   &  90.8 &   90.6 &  91.2  &  88.4 & 88.6 &  91.8 &  91.0 \\
\hline \\
\vspace{0.1cm}
\vphantom & \hspace{0.05cm}  $f_1$ & LPS &   90.2 &  92.8   &  92.0  &   91.0 &  91.6 &  92.4 &  92.4  &  92.6 &  90.2 \\
\vspace{0.1cm}
\vphantom & \hspace{0.05cm}  $f_1$ & LPSMAP &  \hspace{-0.15cm} 90.0 &  92.2  &  91.6  &   91.0 &  91.6 &  92.0 & \hspace{-0.1cm} 91.6 &  92.6 &  89.8 \\
\vspace{0.1cm}
\vphantom & \hspace{0.05cm}  $f_1$ & MGCV &   90.4 &  92.8   & 91.4  &   91.4 &  91.8  &  91.6 &  92.4  &  92.0 &  90.4 \\
\vspace{0.1cm}
\vphantom{} Normal & \hspace{0.05cm}  $f_2$ & LPS &   91.6 &  90.4   &  91.2  &  \hspace{0.05cm} 94.8$^{*}$ &  92.2 & \hspace{0.05cm} 93.6$^{*}$ &  91.2 &  90.0 &  89.4  \\
\vspace{0.1cm}
\vphantom & \hspace{0.05cm}  $f_2$ & LPSMAP &  \hspace{-0.09cm} 91.2 &  89.4  &  90.0  & \hspace{0.08cm}  94.6$^{*}$ &  91.6  &  \hspace{0.08cm}  94.0$^{*}$ & \hspace{-0.11cm} 90.8  &  90.0 &  89.2 \\
\vspace{0.1cm}
\vphantom & \hspace{0.05cm}  $f_2$ & MGCV &  92.0 & 90.4  & 90.8  &  \hspace{0.05cm} 94.4$^{*}$ & 92.0  & \hspace{0.05cm} 93.8$^{*}$ & 92.0  & 91.2 & 89.6 \\
\vspace{0.1cm}
\vphantom & \hspace{0.05cm}  $f_3$ & LPS &  90.4 &  92.0 & \hspace{-0.1cm} 90.6  &   92.4 &  90.8 &  87.4  &  89.4  & \hspace{-0.1cm} 92.6 &  89.6 \\
\vspace{0.1cm}
\vphantom & \hspace{0.05cm}  $f_3$ & LPSMAP &  \hspace{-0.1cm} 90.4 &  92.2   &  \hspace{-0.1cm} 90.4  &   92.2 &  90.6  &  88.0 & \hspace{-0.1cm} 89.0  &  92.4 &  89.2 \\
\vspace{0.1cm}
\vphantom & \hspace{0.05cm}  $f_3$ & MGCV &   89.8 &  92.4   & \hspace{-0.1cm} 91.8 &  \hspace{-0.15cm} 91.6 &  90.0  & 88.8 &  89.8  &  92.4 & 89.6 \\
\hline \\
\vspace{0.1cm}
\vphantom & \hspace{0.05cm}  $f_1$ & LPS &   88.4 & \hspace{0.05cm} 94.0$^{*}$  &  89.2  &  93.0 &  91.0  & \hspace{0.05cm} 96.0$^{*}$ &  91.6  &  90.8 &  88.2 \\
\vspace{0.1cm}
\vphantom & \hspace{0.05cm}  $f_1$ & LPSMAP &  \hspace{-0.11cm} 87.6 &  93.0   &  87.6  &  92.8 & 90.6 &  \hspace{0.05cm} 96.0$^{*}$ & \hspace{-0.11cm} 91.4  &  91.0 &  88.0 \\
\vspace{0.1cm}
\vphantom & \hspace{0.05cm}  $f_1$ & MGCV &   88.6 &  \hspace{0.05cm} 93.8$^{*}$   &  89.4  &   \hspace{0.05cm} 93.4$^{*}$ &  90.6  & \hspace{0.05cm} 96.2$^{*}$ &  93.2  &  91.4 &  89.0 \\
\vspace{0.1cm}
\vphantom{} Binomial & \hspace{0.05cm}  $f_2$ & LPS &   89.8 &  92.6   &  86.8  &   90.8 &  \hspace{0.05cm} 93.6$^{*}$ &  92.8 &  86.8 &  92.0 & \hspace{0.05cm} 84.2$^{*}$  \\
\vspace{0.1cm}
\vphantom & \hspace{0.05cm}  $f_2$ & LPSMAP &  \hspace{-0.1cm} 89.2 &  91.8   &  \hspace{0.08cm} 85.4$^{*}$   &   90.2 &  \hspace{0.08cm} 93.6$^{*}$   &  92.2 &  86.8 &  91.0 &  \hspace{0.05cm} 83.8$^{*}$ \\
\vspace{0.1cm}
\vphantom & \hspace{0.05cm}  $f_2$ & MGCV &  90.0 & \hspace{0.05cm} 94.4$^{*}$   & 87.6  &  92.2 & \hspace{0.05cm} 93.8$^{*}$  & 92.4 & 90.4  & 91.6 & 86.8 \\
\vspace{0.1cm}
\vphantom & \hspace{0.05cm}  $f_3$ & LPS &   87.8 &  91.0 &  87.8  &   90.6 &  90.6 &  86.8  &  87.4  & 92.4 &  90.4 \\
\vspace{0.1cm}
\vphantom & \hspace{0.05cm}  $f_3$ & LPSMAP &  \hspace{-0.1cm} 87.6 &  90.6   &  87.2  &  89.8 &  90.6  &  86.6 & \hspace{0.05cm} 86.2$^{*}$  &  92.2 &  90.0 \\
\vspace{0.1cm}
\vphantom & \hspace{0.05cm}  $f_3$ & MGCV &   88.6 &  91.0   &  89.4  &   91.8 & 89.8  &  89.4 & 89.4  &  92.6 &  90.6 \\
\hline
\end{tabular}
\caption{Effective frequentist coverages of $90\%$ pointwise credible intervals for the functions $f_1, f_2, f_3$ at selected domain points for Poisson, Normal and Binomial data over $S=500$ replications of sample size $n=300$ for the Laplace-P-spline (LPS), the LPS omitting the mixture (LPSMAP) and {\tt{gam}} (MGCV) methods. An asterisk indicates incompatibility with the nominal value.}
\label{table:tab2}
\end{table}

\noindent In the Bernoulli setting where the information content for a given sample size is much smaller than under the other simulation scenarios, all the considered methods exhibit effective frequentist coverages below the nominal value as illustrated in Table \ref{table:tab3} with $n=300$. It corresponds to situations where the estimates of the additive terms provided by LPS(MAP) or {\tt{gam}} can be inaccurate. The pronounced undercoverage in this setting is explained by the poor information conveyed by a binary random variable that translates into oversmoothing of the additive functional components as highlighted in Figure \ref{fig:figure2}. However, as expected, increasing the sample size in the Bernoulli scenario yields frequentist coverage probabilities close to their nominal value (cf. Table \ref{table:tab3} with $n=2000$) both for the LPS(MAP) and {\tt{gam}} methods.\\
\indent Table \ref{table:tab4} reports the effective frequentist coverages of $ 90\%$, $95\%$ and $99\%$ pointwise credible intervals averaged over 200 uniformly distributed values of the covariate on $[-1,1]$ and $S=500$ dataset replications in the Poisson, Normal and Binomial settings. Again, the LPS and LPSMAP methodologies display estimated coverages close to their nominal value in all scenarios. The {\tt{gam}} results show similar performance when coverages are averaged over the covariate support. Note that {\tt{gam}} and LPSMAP rely on a similar approach for selecting the optimal posterior penalty value. Hence, the simulation results presented in this section suggest that our penalty selection scheme is at least as efficient as what is implemented in {\tt{gam}} for estimating the smooth components in the additive part of the model. The simulation results confirm the attractiveness of the Laplace-P-spline model for pointwise and set estimation of the regression parameters in the linear part as well as of the smooth additive components. To enhance the estimation accuracy of our approach in the case of extremely discrete responses such as, for example, Bernoulli data, a possibility is to improve the approximation to the conditional posterior $\widetilde{p}_G(\boldsymbol{\xi} \vert \boldsymbol{\lambda},\mathcal{D})$ by correcting for location and skewness as suggested in \cite{rue2009approximate}. Beyond such extreme binary data configurations, the simple Laplace approximation underlying LPS and LPSMAP suffices for precise inference.


\vspace{0.2cm}

\begin{table}[h!]
\linespread{0.7} \selectfont
\setlength{\tabcolsep}{5pt}
\begin{tabular}{lclccccccccc}
\hline \\
Data &  $f$ &  Method &  -0.95  &  -0.70 &  -0.50 &   -0.20 &  \hspace{0.1cm} 0.00 &   0.20 &   0.50 &   0.70 &  0.95\\
\hline
\\
\vspace{0.1cm}
\vphantom & \hspace{0.05cm}  $f_1$ & LPS & 85.4$^{*}$ &  78.0$^{*}$  &   0.6$^{*}$   &  35.0$^{*}$ &   1.4$^{*}$ & 47.0$^{*}$ &   1.0$^{*}$  &   84.0$^{*}$ &  82.2$^{*}$ \\
\vspace{0.1cm}
\vphantom & \hspace{0.05cm}  $f_1$ & LPSMAP &  86.2$^{*}$ & 78.2$^{*}$  &    0.6$^{*}$  &  25.6$^{*}$ &   0.6$^{*}$ &  46.0$^{*}$ &  0.4$^{*}$  &   84.6$^{*}$ &   82.2$^{*}$ \\
\vspace{0.1cm}
\vphantom & \hspace{0.05cm}  $f_1$ & MGCV & 84.8$^{*}$ &  77.6$^{*}$   &   42.0$^{*}$  &   76.4$^{*}$ &   38.2$^{*}$  &  77.4$^{*}$ &   42.0$^{*}$  &   82.2$^{*}$ &   85.2$^{*}$ \\
\vspace{0.1cm}
\vphantom{} Bernoulli & \hspace{0.05cm}  $f_2$ & LPS & \hspace{-0.3cm} 86.8 &  82.6$^{*}$   &  62.0$^{*}$ & 34.4$^{*}$ &  \hspace{-0.3cm} 86.6 &  52.4$^{*}$ & 58.6$^{*}$ &  \hspace{-0.3cm} 89.6 &  73.0$^{*}$  \\
\vspace{0.1cm}
\hspace{0.1cm} (n=300) & \hspace{0.05cm}  $f_2$ & LPSMAP &   83.2$^{*}$ &    72.8$^{*}$   &  60.6$^{*}$  &  26.8$^{*}$ &  84.2$^{*}$  &   42.6$^{*}$ &  58.0$^{*}$  &  84.8$^{*}$ &   66.6$^{*}$ \\
\vspace{0.1cm}
\vphantom & \hspace{0.05cm}  $f_2$ & MGCV &  \hspace{-0.3cm} 87.8 &  77.0$^{*}$   &  84.8$^{*}$  &   66.0$^{*}$ & \hspace{-0.3cm} 90.0 &   72.2$^{*}$ &  83.8$^{*}$  & 79.6$^{*}$ &  83.2$^{*}$ \\
\vspace{0.1cm}
\vphantom & \hspace{0.05cm}  $f_3$ & LPS &  \hspace{-0.3cm} 88.0 &    80.4$^{*}$ &   2.6$^{*}$  &   1.2$^{*}$ &   96.0$^{*}$ &    1.2$^{*}$  &    2.2$^{*}$  &   71.0$^{*}$ &  77.8$^{*}$ \\
\vspace{0.1cm}
\vphantom & \hspace{0.05cm}  $f_3$ & LPSMAP &   \hspace{-0.3cm} 87.6 &    82.0$^{*}$   &   2.2$^{*}$  &  1.2$^{*}$ &  \hspace{-0.3cm} 92.8  &  1.2$^{*}$ &  1.8$^{*}$  &  65.0$^{*}$ &  62.6$^{*}$ \\
\vspace{0.1cm}
\vphantom & \hspace{0.05cm}  $f_3$ & MGCV &  \hspace{-0.3cm} 87.4 &    84.2$^{*}$   &  52.0$^{*}$  &   51.0$^{*}$ & \hspace{-0.3cm} 90.0 &  48.8$^{*}$ &  49.0$^{*}$  & 83.6$^{*}$ & \hspace{-0.3cm} 86.8 \\
\hline\\
\vspace{0.1cm}
\vphantom & \hspace{0.05cm}  $f_1$ & LPS &  \hspace{-0.3cm} 90.0 & \hspace{-0.3cm} 89.8  &  \hspace{-0.3cm} 87.4   &  94.2$^{*}$ &   \hspace{-0.3cm} 87.4 & \hspace{-0.3cm} 91.8 &  \hspace{-0.3cm} 87.6  &  \hspace{-0.3cm} 89.8 &  \hspace{-0.3cm} 86.6 \\
\vspace{0.1cm}
\vphantom & \hspace{0.05cm}  $f_1$ & LPSMAP &  \hspace{-0.3cm} 89.4 &   \hspace{-0.3cm} 90.2  & \hspace{-0.3cm} 87.0  &   94.0$^{*}$ &  \hspace{-0.3cm} 87.6 &  \hspace{-0.3cm} 92.0 &  \hspace{-0.3cm} 86.8  &  \hspace{-0.3cm} 88.6 &   \hspace{-0.3cm} 86.6 \\
\vspace{0.1cm}
\vphantom & \hspace{0.05cm}  $f_1$ & MGCV &  \hspace{-0.3cm} 89.8 & \hspace{-0.3cm} 91.2   & \hspace{-0.3cm} 90.6  & \hspace{-0.3cm} 93.2 &  \hspace{-0.3cm} 90.8  & \hspace{-0.3cm} 91.6 &  \hspace{-0.3cm} 90.6  &  \hspace{-0.3cm} 89.2 &  \hspace{-0.3cm} 87.8 \\
\vspace{0.1cm}
\vphantom{} Bernoulli & \hspace{0.05cm}  $f_2$ & LPS &  \hspace{-0.3cm}  88.8 &  \hspace{-0.3cm} 90.8 & \hspace{-0.3cm} 87.0  &   \hspace{-0.3cm} 89.8 &  \hspace{-0.3cm} 93.0 &  \hspace{-0.3cm} 90.8 &  \hspace{-0.3cm} 86.6 & \hspace{-0.3cm} 91.2 & \hspace{-0.3cm} 86.8  \\
\vspace{0.1cm}
\hspace{0.05cm} (n=2000) & \hspace{0.05cm}  $f_2$ & LPSMAP &   \hspace{-0.3cm} 87.6 &   \hspace{-0.3cm} 90.6 &  86.2$^{*}$   & \hspace{-0.3cm}  89.0 &  \hspace{-0.3cm} 92.6  &  \hspace{-0.3cm} 90.6 & \hspace{-0.3cm}  86.6  & \hspace{-0.3cm} 90.4 & \hspace{-0.3cm} 86.6 \\
\vspace{0.1cm}
\vphantom & \hspace{0.05cm}  $f_2$ & MGCV & \hspace{-0.3cm} 89.2 & \hspace{-0.3cm} 91.8  & \hspace{-0.3cm} 88.8  &  \hspace{-0.3cm} 90.6 & \hspace{-0.3cm} 93.2 &  \hspace{-0.3cm} 91.4 & \hspace{-0.3cm} 90.0  & \hspace{-0.3cm} 90.6 & \hspace{-0.3cm} 91.2 \\
\vspace{0.1cm}
\vphantom & \hspace{0.05cm}  $f_3$ & LPS &  \hspace{-0.3cm} 90.2 &   \hspace{-0.3cm} 88.2 &  86.0$^{*}$  & \hspace{-0.3cm} 87.6 &  \hspace{-0.3cm} 93.2 &  84.8$^{*}$  &  84.4$^{*}$  & \hspace{-0.3cm} 89.2 &  \hspace{-0.3cm} 91.2 \\
\vspace{0.1cm}
\vphantom & \hspace{0.05cm}  $f_3$ & LPSMAP & \hspace{-0.3cm} 90.4 &   \hspace{-0.3cm} 87.8  &  84.8$^{*}$  & \hspace{-0.3cm} 87.2 &  \hspace{-0.3cm} 93.0  &  83.8$^{*}$ & 83.0$^{*}$  & \hspace{-0.3cm} 89.2 &  \hspace{-0.3cm} 90.6 \\
\vspace{0.1cm}
\vphantom & \hspace{0.05cm}  $f_3$ & MGCV & \hspace{-0.3cm} 90.8 &   \hspace{-0.3cm} 88.6  & \hspace{-0.3cm} 89.6  &  \hspace{-0.3cm} 91.4 & \hspace{-0.3cm} 92.2 & \hspace{-0.3cm} 88.6 & \hspace{-0.3cm} 87.0  &  \hspace{-0.3cm} 90.2 & \hspace{-0.3cm} 91.2 \\
\hline
\end{tabular}
\caption{Effective frequentist coverages of $90\%$ pointwise credible intervals for the functions $f_1, f_2, f_3$ at selected domain points for Bernoulli data over $S=500$ replications of sample size $n=300$ and $n=2000$ for the Laplace-P-spline (LPS), the LPS omitting the mixture (LPSMAP) and {\tt{gam}} (MGCV) methods. An asterisk indicates incompatibility with the nominal value.}
\label{table:tab3}
\end{table}

\newpage 


\begin{table}[h!]
\linespread{0.6} \selectfont
\setlength{\tabcolsep}{4pt}
\begin{tabular}{llccccccccccc}
\hline \\
\vphantom & \vphantom & \vphantom & $90\%$ & \vphantom  & \vphantom  & \vphantom & $95\%$ & \vphantom & \vphantom & \vphantom & $99\%$ &\phantom{k}  \\
\cline{3-5}   \cline{7-9} \cline{11-13} \vspace{0.1cm} \hspace{0.1cm}\\
Data & \hspace{0.15cm} Method & \hspace{0.25cm} $f_1$  &  \hspace{0.25cm} $f_2$ &  \hspace{0.25cm} $f_3$ & & \hspace{0.25cm} $f_1$ &  \hspace{0.25cm} $f_2$ &  \hspace{0.25cm} $f_3$ & & \hspace{0.25cm} $f_1$ &  \hspace{0.25cm} $f_2$ & \hspace{0.25cm} $f_3$\\
\hline \\
\vspace{0.1cm}
Poisson & \hspace{0.15cm} LPS &  \hspace{0.15cm} 87.6 & \hspace{0.15cm} 87.0   & \hspace{0.15cm} 89.1  &  & \hspace{0.15cm} 93.0 &   \hspace{0.1cm} 92.6  & \hspace{0.1cm} 94.4 & & \hspace{0.1cm} 98.0  & \hspace{0.1cm} 98.1 & \hspace{0.1cm} 98.9 \\
\vspace{0.1cm}
\vphantom & \hspace{0.15cm} LPSMAP &  \hspace{0.15cm} 86.7 & \hspace{0.15cm} 85.6  & \hspace{0.15cm} 88.7  &  & \hspace{0.15cm} 92.4 &   \hspace{0.1cm} 91.6  & \hspace{0.1cm} 94.0 & & \hspace{0.1cm} 97.7  & \hspace{0.1cm} 97.4 & \hspace{0.1cm} 98.7 \\
\vspace{0.1cm}
\vphantom  & \hspace{0.15cm} MGCV &  \hspace{0.15cm} 89.8 & \hspace{0.15cm} 89.6  & \hspace{0.15cm} 90.3  & & \hspace{0.15cm} 94.4 & \hspace{0.1cm} 94.4  & \hspace{0.1cm} 95.1 & & \hspace{0.1cm} 98.8  & \hspace{0.1cm} 98.7 & \hspace{0.1cm} 99.1 \\
\vspace{0.1cm}
Normal & \hspace{0.15cm} LPS &  \hspace{0.15cm} 90.8 & \hspace{0.15cm} 91.1   & \hspace{0.15cm} 91.0  & &  \hspace{0.15cm} 95.6 & \hspace{0.1cm} 95.8 & \hspace{0.1cm} 95.8 & & \hspace{0.1cm} 99.2 & \hspace{0.1cm} 99.0 & \hspace{0.1cm} 99.3  \\
\vspace{0.1cm}
\vphantom & \hspace{0.15cm} LPSMAP &  \hspace{0.15cm} 90.5 & \hspace{0.15cm} 90.7   & \hspace{0.15cm} 90.9  &  & \hspace{0.15cm} 95.4 &   \hspace{0.1cm} 95.4  & \hspace{0.1cm} 95.6 & & \hspace{0.1cm} 99.2  & \hspace{0.1cm} 99.0 & \hspace{0.1cm} 99.3 \\
\vspace{0.1cm}
\vphantom  & \hspace{0.15cm} MGCV &  \hspace{0.15cm} 91.1 & \hspace{0.15cm} 91.5  & \hspace{0.15cm} 91.2 & &  \hspace{0.15cm} 95.8 & \hspace{0.1cm} 95.8  & \hspace{0.1cm} 95.8 & & \hspace{0.1cm} 99.3  & \hspace{0.1cm} 99.1 & \hspace{0.1cm} 99.3 \\
\vspace{0.1cm}
Binomial &  \hspace{0.15cm} LPS &  \hspace{0.15cm} 90.2 & \hspace{0.15cm} 89.3 & \hspace{0.15cm} 90.3  & & \hspace{0.15cm} 95.0 & \hspace{0.1cm} 94.5 & \hspace{0.1cm} 95.3  & & \hspace{0.1cm} 98.8  & \hspace{0.1cm} 98.8 & \hspace{0.1cm} 99.1 \\
\vspace{0.1cm}
\vphantom & \hspace{0.15cm} LPSMAP &  \hspace{0.15cm} 89.9 & \hspace{0.15cm} 88.8  & \hspace{0.15cm} 90.1  &  & \hspace{0.15cm} 94.7 &   \hspace{0.1cm} 94.1  & \hspace{0.1cm} 95.1 & & \hspace{0.1cm} 98.7  & \hspace{0.1cm} 98.6 & \hspace{0.1cm} 99.1 \\
\vspace{0.1cm}
\vphantom  & \hspace{0.15cm} MGCV &  \hspace{0.15cm} 91.2 & \hspace{0.15cm} 90.2  & \hspace{0.15cm} 90.9  & &  \hspace{0.15cm} 95.4 & \hspace{0.1cm} 95.1  & \hspace{0.1cm} 95.6 & & \hspace{0.1cm} 99.0  & \hspace{0.1cm} 98.9 & \hspace{0.1cm} 99.2 \\
\hline
\end{tabular}
\caption{Effective frequentist coverages of $90\%$, $95\%$ and $99\%$ pointwise credible intervals averaged over 200 uniformly distributed values of the covariate $x$ in $[-1,1]$ for Poisson, Normal and Binomial data with $S=500$ replications of sample size $n=300$ for the Laplace-P-spline (LPS), the LPS omitting the mixture (LPSMAP) and {\tt{gam}} (MGCV) methods.}
\label{table:tab4}
\end{table}


\begin{figure}[h!] 
\centering
\includegraphics[height=5cm, width=16.5cm]{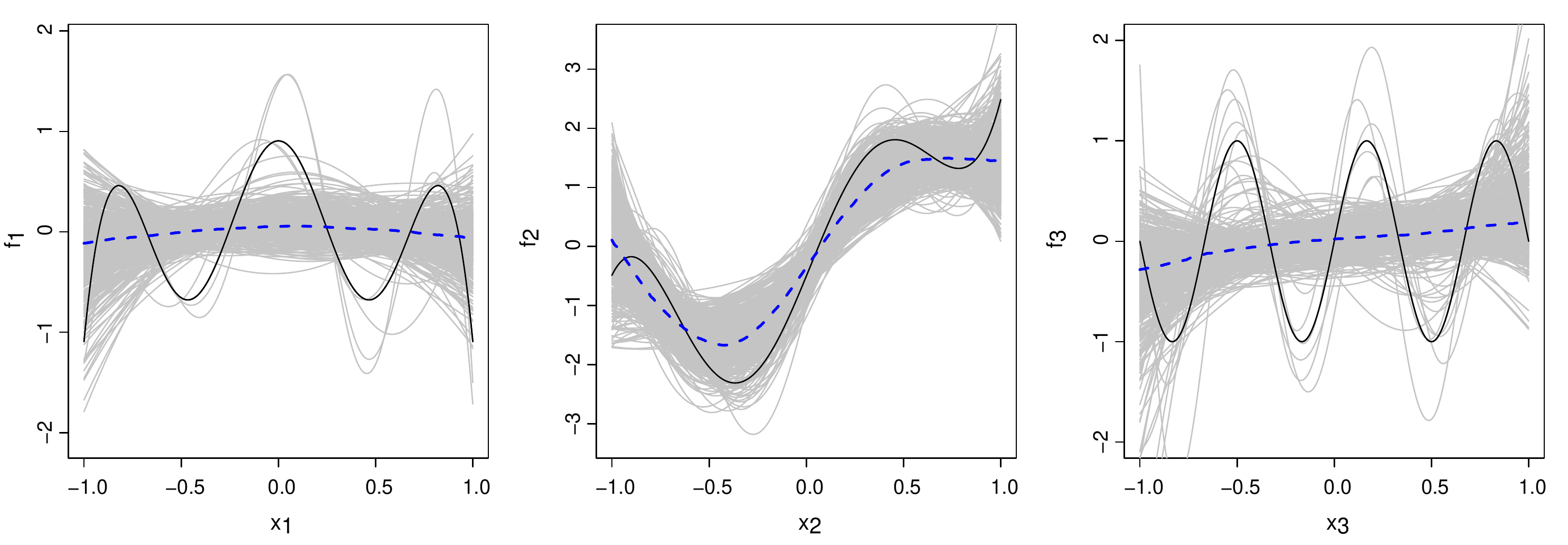}
\caption{Estimation of smooth additive terms (gray curves) for $S=500$ dataset replications of size $n=300$ in the Bernoulli scenario with LPS. The dashed line is the pointwise median of the gray curves and the black curves are the target functions.}
\label{fig:figure2}
\end{figure}

\vspace{-0.7cm}

\subsection{Computational costs}

\noindent To illustrate the computational behavior of LPS and LPSMAP against sample size for fixed dimension $q=3$, we consider an increasing sequence of sample sizes from $n=200$ to $n=3000$ in steps of 200 and for each considered sample size compute the average wall clock time (elapsed real time) in seconds with the {\tt{proc.time()}} function in {\bf{R}} over $10$ different samples. In Figure \ref{fig:figure3} (a) the elapsed time to estimate the GAM model with LPS and LPSMAP is plotted against sample size to depict the involved computational resources. Both curves show an exponential increase with sample size. LPSMAP is faster than LPS as it does not require a grid construction to explore the support of the marginal posterior of the penalty parameters, but rather fix them at their posterior mode. Figure \ref{fig:figure3} (b) highlights the computational time of LPSMAP against sample size $n$ on a log-log scale.


\begin{figure}[h!] 
\centering
\includegraphics[height=6.3cm, width=16.4cm]{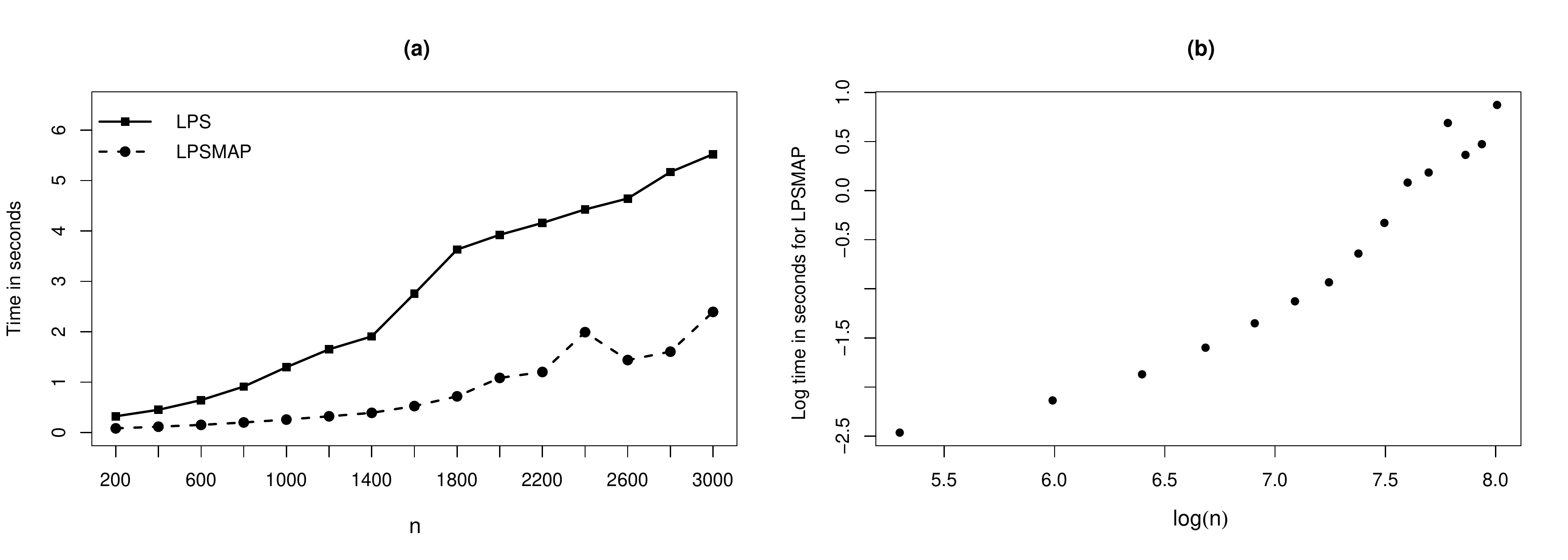}
\caption{(a) Real elapsed time in seconds as a function of sample size for LPS and LPSMAP. (b) Computational time of LPSMAP against sample size on a log-log scale.}
\label{fig:figure3}
\end{figure}

\newpage 

\subsection{Simulation study with more additive terms.}

\noindent A large number $q$ of smooth functions in the additive predictor implies an increased computational burden. Algorithm 1 suggests to prefer independence sampling over a grid construction to explore the marginal posterior of the penalty parameters when $q>4$, see Section 2.5.2 for details. To illustrate how the Laplace-P-spline model performs with a larger number of smooth functions,  we simulate $S=500$ datasets of size $n=300$ and a Markov chain sample of size $500$ for each replicate with the following additive terms: 

\vspace{-0.3cm}

\begin{eqnarray}
f_1(x_1)&=&0.5 (2x_1^5+3x_1^2+\cos(3\pi x_1)-1), \nonumber \\
f_2(x_2)&=&1.3x_2^5+\sin(4x_2)+0.75x_2^2-0.25, \nonumber \\
f_3(x_3)&=&\sin(4 \pi x_3), \nonumber \\
f_4(x_4)&=&\exp(-x_4^3) \sin(2 \pi x_4^2)-0.1, \nonumber \\
f_5(x_5)&=&0.8 x_5^2 (x_5^3+2 \exp(-3 x_5^4+\log(2x_5+\pi)))-0.65, \nonumber \\
f_6(x_6)&=&1.5\ \big(0.1\sin(2 \pi x_6)+0.2\cos(2 \pi x_6)+0.3\sin^2(2 \pi x_6) \nonumber \\
&&+0.4\cos^3(2 \pi x_6)+0.5\sin^3(2 \pi x_6) \big)-0.22. \nonumber 
\end{eqnarray}

\vspace{0.1cm}

\noindent There are three additional covariates specified as in Section \ref{sec:6.1} with regression coefficients $\beta_0=-1.20$, $\beta_1=0.50$, $\beta_2=-0.40$ and $\beta_3=0.70$. The covariates of the smooth functions are drawn independently from the Uniform distribution on the domain $[-1,1]$. Each smooth function is modeled using a linear combination of $15$ cubic B-splines associated to equidistant knots on $[-1,1]$ and a third order penalty to control smoothness. Two scenarios are considered for the generating process of the response, namely  (1) a Gaussian model $y_i \sim \mathcal{N}(\mu_i, \sigma^2=0.5)$ and (2) a Binomial model $y_i \sim \text{Bin}(20, p_i)$, with $p_i$ the success probability and a logit link function. Table \ref{table:tab5} shows the simulation results of the Laplace-P-spline approach combined with MCMC (cf. Section 2.5.2). The estimation results obtained with the {\tt{gam}} function from the {\ttfamily{mgcv}} package are shown in parenthesis.\\
\indent Estimated biases shown in Table \ref{table:tab5} are almost similar for the two different approaches and nearly equal to zero in the considered data scenarios. In addition, the reported coverage probabilities are close to their corresponding nominal value and analogous results appear for the ESE and RMSE with the LPS and {\tt{mgcv}} algorithms. Figure \ref{fig:figure4} illustrates the estimation results for the six additive smooth terms with the proposed Laplace-P-spline methodology in the Binomial case.

\vspace{-0.4cm}


\begin{center}
\begin{table}[h!]
\linespread{0.7} \selectfont
\setlength{\tabcolsep}{2.5pt}
\begin{tabular}{lcccccc}
\hline \\
Data &  Parameters &  Bias  &  CP$_{90\%}$ &  CP$_{95\%}$ & ESE &  RMSE \\
\hline \\
\vspace{0.1cm}
\phantom{k} & $\beta_1=\phantom{-}0.50$ & \hspace{0.2cm} \phantom{-}0.001 (\phantom{-}0.001) & \hspace{0.1cm} 87.8 (87.4) & \hspace{0.1cm} 94.0 (94.6) & 0.096 (0.095) & 0.096 (0.095) \\
\vspace{0.1cm}
Normal  & $\beta_2=-0.40$ & \hspace{0.2cm} \phantom{-}0.003 (\phantom{-}0.003) & \hspace{0.1cm} 86.8 (87.4) & \hspace{0.1cm} 94.8 (95.0) & 0.047 (0.047) & 0.047 (0.047) \\
\vspace{0.1cm}
\phantom{k} & $\beta_3=\phantom{-}0.70$ & \hspace{0.2cm} \phantom{-}0.003 (\phantom{-}0.003) & \hspace{0.1cm} 86.2 (86.8) & \hspace{0.1cm} 93.2 (92.2) & 0.049 (0.049) & 0.049 (0.049) \\
\hline 
\\
\vspace{0.1cm}
\phantom{k} & $\beta_1=\phantom{-}0.50$ & \hspace{0.2cm} -0.007 (-0.003) & \hspace{0.1cm} 89.6 (89.6) & \hspace{0.1cm} 93.4 (94.0) & 0.078 (0.078) & 0.079 (0.078) \\
\vspace{0.1cm}
Binomial & $\beta_2=-0.40$ & \hspace{0.25cm} 0.003 (\phantom{-}0.000) & \hspace{0.1cm} 88.8 (89.6) & \hspace{0.1cm} 94.4 (94.4) & 0.041 (0.041) & 0.041 (0.041) \\
\vspace{0.1cm}
\phantom{k} & $\beta_3=\phantom{-}0.70$ & \hspace{0.2cm} -0.009 (-0.003) & \hspace{0.1cm} 87.8 (88.2) & \hspace{0.1cm} 94.2 (95.0) & 0.043 (0.043) & 0.044 (0.043) \\
\hline
\end{tabular}
\caption{Simulation results for $S=500$ replicates of sample size $n=300$ for Normal and Binomial data when independence sampling is used to draw samples from $\tilde{p}(\mathbf{v} \vert \mathcal{D})$. The values in parentheses are estimation results from the {\tt{gam}} function.}
\label{table:tab5}
\end{table}
\end{center}

\vspace{-1cm}

\noindent For each graph, there are $S=500$ gray curves representing the estimates of the corresponding unknown smooth function (black) entering the additive predictor. The dashed curve represents the pointwise median of the $500$ estimated curves. For each smooth term, the observed estimates are close to the target, even with highly oscillating functions (e.g. $f_3$ and $f_6$). For function $f_6$, small bumps arising near main curvatures can be better captured by increasing the number of B-splines in the basis.

\vspace{-0.3cm}


\begin{figure}[h!] 
\centering
\includegraphics[height=9.2cm, width=16.5cm]{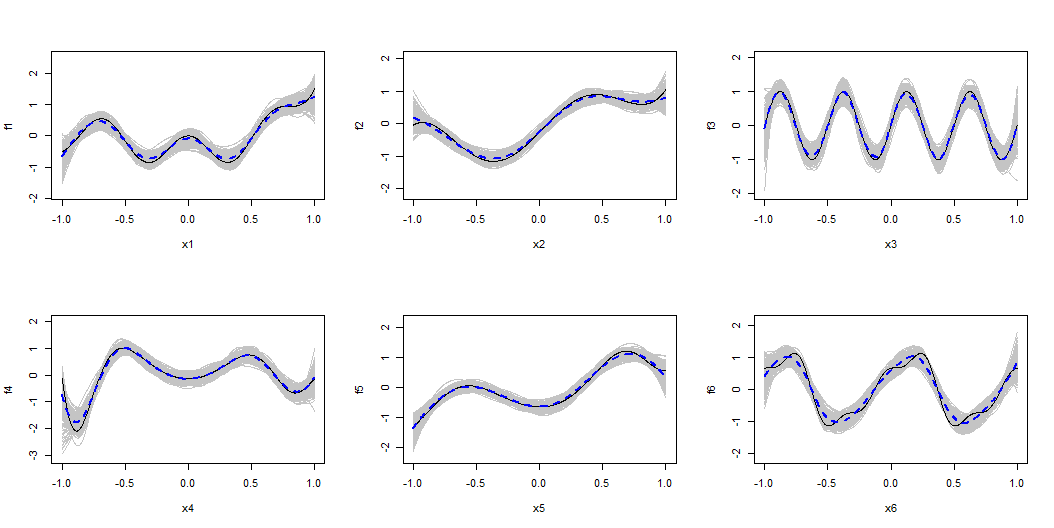}
\caption{Estimation of smooth additive terms $f_1,\dots,f_6$ (gray curves) for $S=500$ dataset replications of size $n=300$ in the Binomial scenario. The dashed line is the pointwise median of the gray curves.}
\label{fig:figure4}
\end{figure}

\newpage 

\noindent With $q=6$, our LPS methodology coupled with MCMC (LPS-MCMC) requires (to build a chain of length $500$) on average 4.70 seconds for a dataset of size $n=300$. In Table \ref{table:tab6}, we provide computation times of the LPS-MCMC algorithm to estimate the GAM for different dimensions $q$ and sample sizes. As expected the computation time increases with $q$ and $n$. Figure 5 gives an overview of the average computational times required to estimate the GAM with the LPS and LPS-MCMC algorithms for an increasing number of additive terms. When $q\leq4$ the LPS approach is faster, but in larger dimensions the LPS-MCMC algorithm (with an independence sample of length 500) requires less computational budget than the grid construction in LPS.

\vspace{-0.1cm}


\begin{center}
\begin{table}[h!]
\linespread{0.7} \selectfont
\setlength{\tabcolsep}{15pt}
\begin{tabular}{ccccccc}
\hline \\
Dimension &  & Average computation time (in seconds) &  \\
\\
&  $n=300$ &    $n=1000$   &   $n=3000$ \\
\hline \\
\vspace{0.1cm}
$q=1$ & 1.86 & 2.78 & 7.00  \\
\vspace{0.1cm}
$q=2$   & 2.10 & 3.46 & 11.60 \\
\vspace{0.1cm}
$q=3$  & 2.51 & 4.66 & 15.09 \\
\vspace{0.1cm}
$q=4$  & 3.04 & 6.53 & 21.04 \\
\vspace{0.1cm}
$q=5$  & 3.82 & 8.83 & 27.55  \\
\vspace{0.1cm}
$q=6$  & 4.70 & 11.46 & 36.08 \\
\hline 
\end{tabular}
\caption{Average computation time (in seconds) of the LPS-MCMC algorithm over $S=20$ samples of size $n \in \{300,1000,3000\}$ for different dimensions $q \in \{1,2,3,4,5,6\}$.}
\label{table:tab6}
\end{table}
\end{center}

\vspace{-0.9cm}


\begin{figure}[h!] 
\centering
\includegraphics[height=7.3cm, width=13cm]{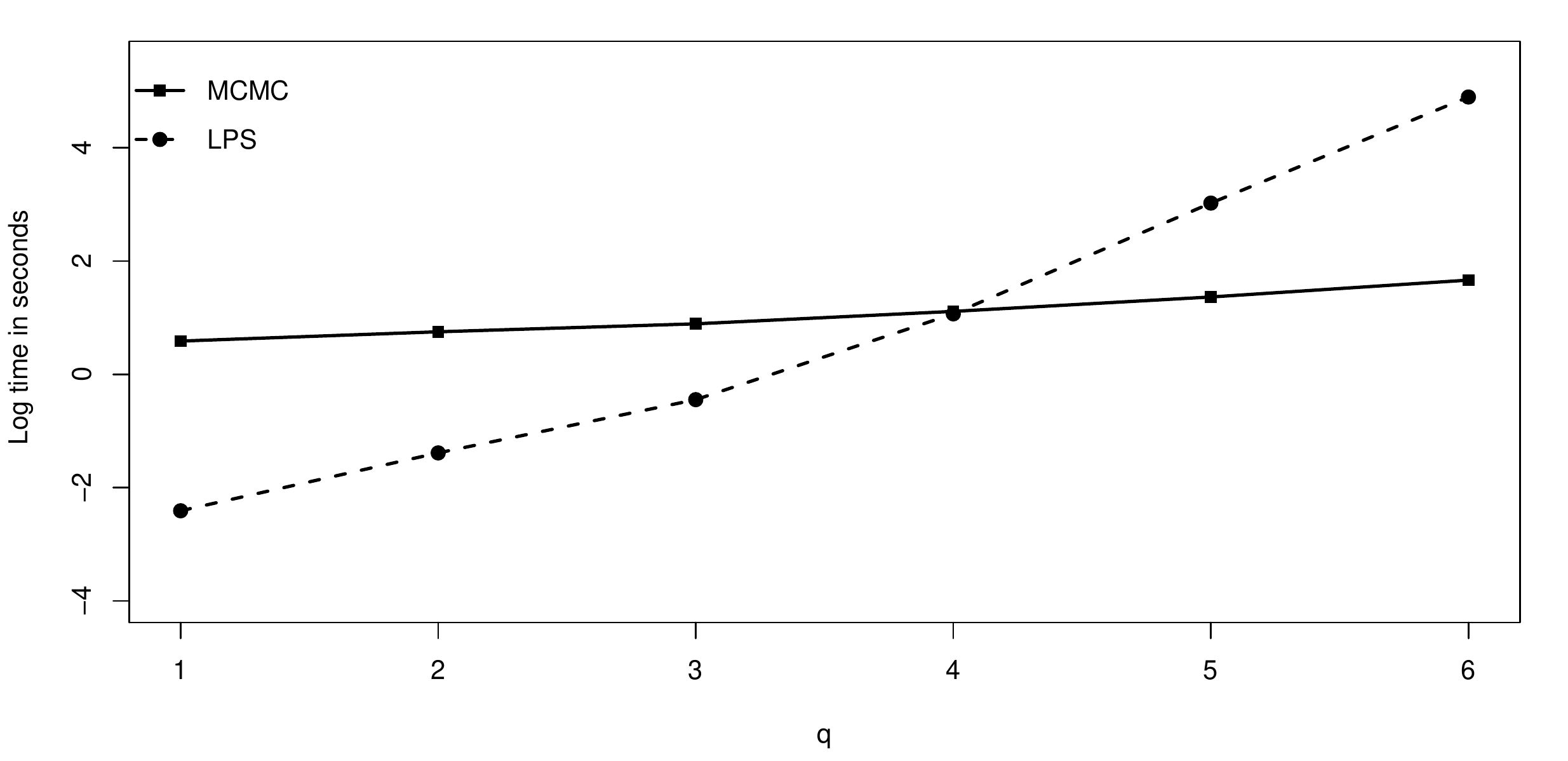}
\caption{Logarithm of the average computation time (in seconds) of LPS (dahsed) and LPS-MCMC (solid) over $S=20$ samples of size $n=300$ and dimensions $q \in \{1,2,3,4,5,6\}$.}
\label{fig:figure5}
\end{figure}

\newpage

\section{Applications}
\label{sec:sec7}

\subsection{Model for the number of doctor visits}

\noindent We apply our Laplace-P-spline model in the context of a health-care  study on Medicaid eligibles. The data are from the 1986 Medicaid Consumer Survey sponsored by the Health Care Financing Administration in the USA. This Medicaid database has first been studied by \cite{gurmu1997semi} in the framework of a semi-parametric hurdle model and later by \cite{sapra2013generalized} as an econometric application of generalized additive models using the $\tt{mgcv}$ package in {\bf{R}}. Our analysis will focus on a sample of $n=485$ adults who meet the requirement for eligibility in the Aid to Families with Dependent Children (AFDC) program. The response variable is the number of doctor visits (office/clinic and health center) over a period of 120 days. The explanatory variables included in the linear part of the GAM are \textit{Children} (Total number of children in the household), \textit{Race} (0=other; 1=white) and \textit{Maritalstatus} (0=other; 1=married). The variables modeled in the smooth nonlinear part are taken to be \textit{Age}, the household annual \textit{Income} (in US dollars), a variable measuring the ease of \textit{Access} to health services with values in the interval (0=low access; 100=high access) and the first principal component built from three health-status variables (functional limitations, acute conditions, chronic conditions) denoted by \textit{PC1} with larger positive numbers meaning poorer health. Descriptive statistics of these variables are detailed in \cite{gurmu1997semi}. The GAM model with a Poisson conditional distribution $\text{Poisson}(\mu_i)\ (i=1,...,n)$ for the number of doctor visits can be written as follows:

\vspace{-0.3cm}

\begin{eqnarray}
g(\mu_i)&=&\beta_0+\beta_1 \textit{Children}_i+\beta_2 \textit{Race}_i +\beta_3 \textit{Maritalstatus}_i \nonumber \\
&&+f_1(\textit{Age}_i)+f_2(\textit{Income}_i)+f_3(\textit{Access}_i)+f_4(PC1_i),\ \ i=1,\dots,n, \nonumber
\end{eqnarray}

\vspace{0.2cm}

\noindent where $g(\cdot)$ is the log-link and the smooth functions $f_j$ are modeled using a linear combination of 15 cubic B-splines penalized by a third order penalty. The B-spline bases are defined over the domain $[x_{j,\text{min}},x_{j,\text{max}}]$, where $x_{j,\text{min}}$ ($x_{j,\text{max}}$) is the minimum (maximum) of the covariate values on which $f_j$ is defined. Given the moderate number of additive terms ($q=4$), the posterior penalty space is explored via the grid strategy of Section 2.5.1.\\
\noindent Table \ref{table:tab7} summarizes the estimation results for the parametric linear part of the GAM. The results highlight a negative and significant relationship between the number of children in a household and the (mean) number of doctor visits. 


\begin{center}
\begin{table}[h!]
\linespread{0.8}\selectfont
\setlength{\tabcolsep}{26pt}
\begin{tabular}{lccc}
\hline \\
\textbf{Parameters} & \textbf{Estimates} &  \textbf{CI 90\%} &  \textbf{sd}$_{post}$ \\
\hline \\
$\beta_1$ (\textit{Children}) &  -0.179 & [-0.239; -0.122] &  0.036\\
$\beta_2$ (\textit{Race}) &  -0.127 & [-0.263; \phantom{-}0.005] &  0.081\\ 
$\beta_3$ (\textit{Maritalstatus}) &  -0.234 & [-0.431; -0.043] &  0.118 \\
\hline
\end{tabular}
\caption{Estimation results for the parametric linear part of the GAM. The second column is the parameter estimate, the third column gives the associated $90\%$ credible interval and the last column is the posterior standard deviation.}
\label{table:tab7}
\end{table}
\end{center}

\newpage 

\noindent  The demographic variable \textit{Race} has a non-significant effect on the the mean response, while a negative and significant relationship between \textit{Maritalstatus} and the (mean) number of doctor visits is observed. Figure \ref{fig:figure6} displays the estimated smooth functions (solid curves) and the associated $95\%$ approximate pointwise credible intervals (gray surfaces). As in \cite{gurmu1997semi}, we observe a concave relationship between the mean response and Age with a peak in the average number of visits arising around \textit{Age}=28. As most of the AFDC beneficiaries are women the concave pattern of Age may be explained by pregnancy-related visits during fertile periods and less frequent visits in later periods of life. The socio-economic variable \textit{Income} exhibits no significant effect on the mean number of doctor visits when \textit{Income} is below \textit{Income}$^*=$10,000\ $\$$. Hence an increase in income for poor households with an annual income below \textit{Income}$^*$ is (on average) not reflected by an increase in the number of doctor visits. However, when the annual income goes above \textit{Income}$^*$ individuals tend to care more about their health and the (average) number of medical visits increases.


\begin{figure}[h!] 
\centering
\includegraphics[height=7.9cm, width=15.2cm]{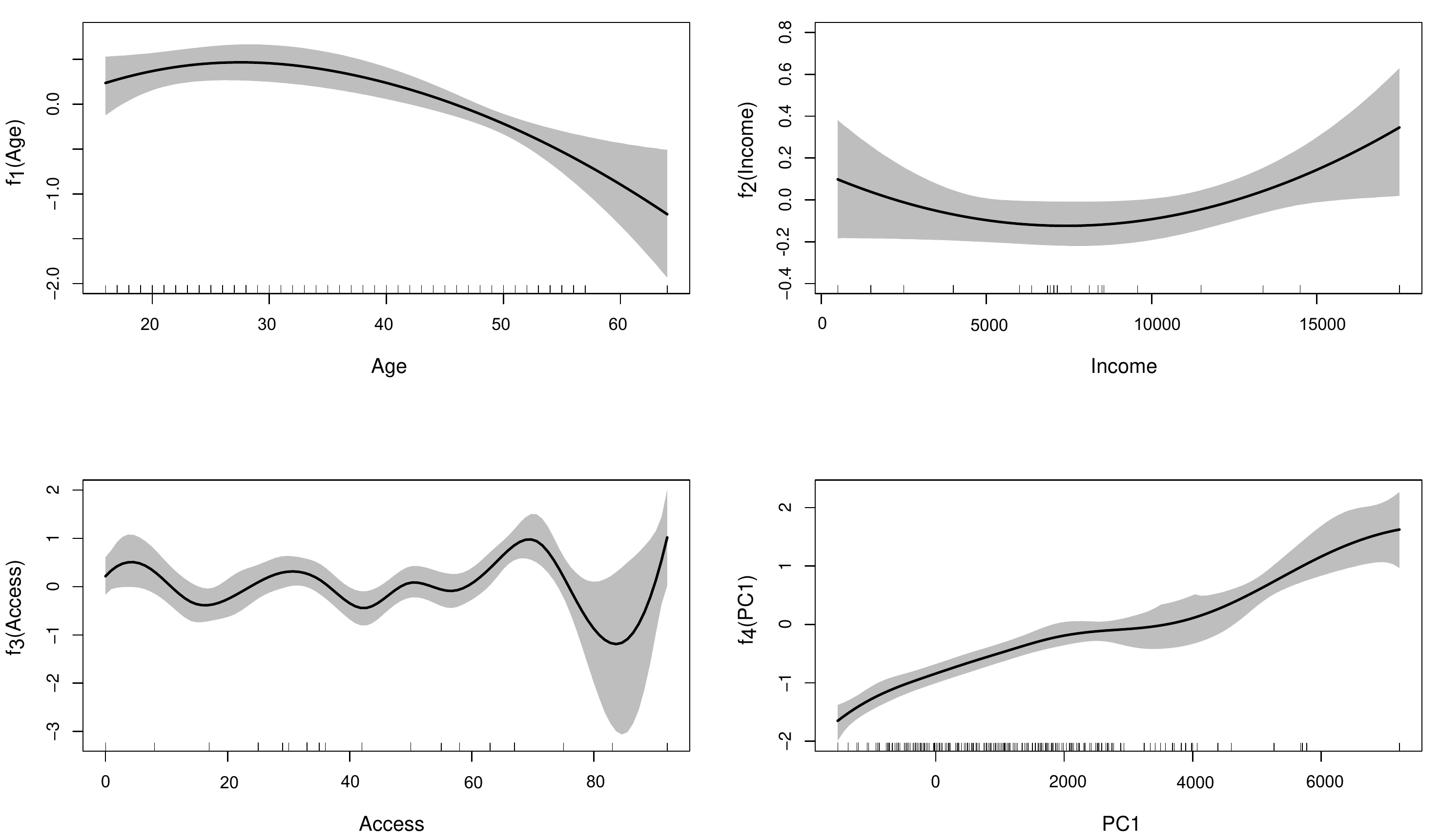}
\caption{Estimated smooth functions (solid curve) and $95\%$ approximate pointwise credible intervals (gray surface)  for variables \textit{Age}, \textit{Income}, \textit{Access} and \textit{PC1}.}
\label{fig:figure6}
\end{figure}

\noindent Furthermore for the variable \textit{Access}, we observe a strong oscillation of the mean response around a linear trend in the domain $[0,70]$, suggesting that for low to moderate health service availability, the mean number of doctor visits remains stable. With regard to health-status variables gathered in \textit{PC1} the results are as expected. Indeed, we observe a clear upward trend, i.e. the average number of medical visits increases with poorer health conditions. 

\subsection{Nutritional study}

In a second application, we implement our methodology to analyze data from a nutritional epidemiology study. More thoroughly, we are interested in modeling the relationship between the plasma beta-carotene level and several explanatory variables related to individual factors and dietary characteristics. Human cells are driven by an important dynamic called the oxidation process, an energy delivery mechanism that is crucial for a proper functioning at the cellular level. By-products of the oxidation process are molecules known as free radicals. An imbalance between free radicals and antioxidant defenses generates oxidative stress which in turn triggers carcinogenesis. Beta-carotene is an antioxidant acting as a free radical scavenger and has been shown to prevent various cancer types and other diseases (\citealp{comstock1992serum}; \citealp{rimm1993vitamin} and \citealp{zhang1999dietary}).\\
\indent The dataset provided by \cite{stukel2008determinants} on plasma beta-carotene levels has $n=314$ observations on 14 variables. Factors influencing beta-carotene plasma concentration levels have been studied by \cite{nierenberg1989determinants}, who found that beta-carotene level had a positive relationship with dietary beta-carotene consumption and tends to be larger for females, whereas a negative relation appeared with current smoker status. The dataset was also analyzed by \cite{liu2011estimation} who develop a variable selection procedure to identify the significant linear components in a semiparametric additive partial linear model. The Laplace-P-spline model is implemented on the \cite{stukel2008determinants} data to study the relationship between the logarithm of beta-carotene plasma level (in ng/ml) and various explanatory variables retained as significant by the analysis in \cite{liu2011estimation}. The linear part of the additive model will include the \textit{BMI} or Quetelet index (weight/height$^2$), the dietary beta-carotene consumption (\textit{Betadiet}) (in mg/day), \textit{Gender} (0=Male; 1=Female), a binary indicator \textit{Smoking} status (0=non smoker; 1=current smoker) and the covariates \textit{Fiber} and \textit{Fat} indicating the hectograms of fiber and fat respectively consumed on a daily basis. The nonlinear part of the model will encompass the variables \textit{Age} (in years) and the log of \textit{Cholesterol} consumption (in mg/day). To summarize, the GAM model with an identity link is given by $y_i=\log(\textit{Betaplasma}_i) \sim \mathcal{N}(\mu_i,s^2)$ where $s^2=0.559$ is the empirical variance of the response and the mean is modeled as:

\vspace{-0.3cm}

\begin{eqnarray}
\mu_i&=&\beta_0+\beta_1 \textit{BMI}_i+ \beta_2 \textit{Betadiet}_i +\beta_3 \textit{Gender}_i+ \beta_4 \textit{Smoking}_i+ \beta_5 \textit{Fiber}+ \beta_6 \textit{Fat} \nonumber \\
&&+f_1(\textit{Age}_i)+f_2(\log(\textit{Cholesterol}_i)), \ i=1,\dots,n. \nonumber
\end{eqnarray}

\vspace{0.1cm}

\noindent In Table 8, we report the estimation results of the linear part.  All variables are significant, except \textit{Betadiet}. There is a negative association between \textit{BMI} and the mean log plasma beta-carotene level meaning that for a fixed height, individuals with lower weight tend to have (on average) higher plasma beta-carotene concentrations. As in \cite{nierenberg1989determinants}, we find that females and non-smokers tend to have a significantly larger beta-response level. A possible explanation is that smoke actually deteriorates beta-carotene molecules through an oxidation process. Finally, fiber consumption increases the mean plasma beta-carotene level, with the consumption of vegetables on a daily basis helping to maintain antioxidants at a high level, while a high-fat diet tends to have a negative effect on the mean response. \\
\indent Figure \ref{figure:fig7} highlights the estimated smooth functions for \textit{Age} and $\log$ \textit{Cholesterol}. For variable \textit{Age} the shape of the estimated function is similar to what is observed in \cite{liu2011estimation}. There is a positive association with the mean response when \textit{Age} is smaller than 45 years or greater than 65 years. On the other hand, the relation of the mean response to the log-cholesterol level does not appear significant.

\newpage 


\begin{center}
\begin{table}[h!]
\linespread{0.9}\selectfont\setlength{\tabcolsep}{28pt}
\begin{tabular}{lccc}
\hline \\
\textbf{Parameters} & \textbf{Estimates} &  \textbf{CI 90\%} &  \textbf{sd}$_{post}$ \\
\hline \\
$\beta_1$ (\textit{BMI}) & -0.034 & [-0.046; -0.022] &  0.007\\
$\beta_2$ (\textit{Betadiet}) &  \phantom{-}0.047 & [-0.009; \phantom{-}0.101] &  0.033\\ 
$\beta_3$ (\textit{Gender}) &  \phantom{-}0.300 & [\phantom{-}0.076; \phantom{-}0.520] &  0.135 \\
$\beta_4$ (\textit{Smoking}) &  -0.301 & [-0.515; -0.093] &  0.128 \\
$\beta_5$ (\textit{Fiber}) &  \phantom{-}2.396 & [\phantom{-}0.804; \phantom{-}3.938] &  0.956 \\
$\beta_6$ (\textit{Fat}) & -0.245 & [-0.493; -0.003] &  0.149 \\
\hline
\end{tabular}
\caption{Estimation results for the parametric linear part of the GAM for the nutritional study. The second column is the parameter estimate, the third column gives the associated $90\%$ credible interval and the last column is the posterior standard deviation.}
\end{table}
\label{table:tab8}
\end{center}


\begin{figure}[h!] 
\centering
\includegraphics[height=5.5cm, width=15.3cm]{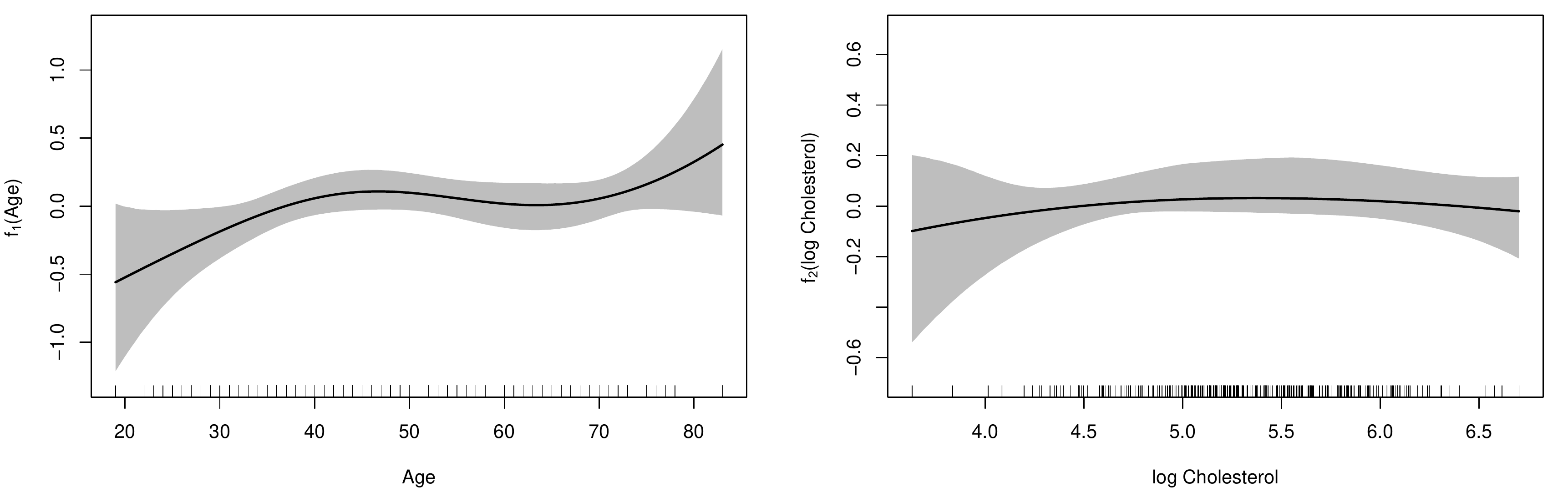}
\caption{Estimated smooth functions (solid curve) and $95\%$ approximate pointwise credible intervals (gray surface) for variables \textit{Age} and $\log$(\textit{Cholesterol}) of the nutritional study dataset.}
\label{figure:fig7}
\end{figure}

\vspace{-0.4cm}

\section{Concluding remarks}
\label{sec:conc}

In this article, we have put forward a new methodology for approximate Bayesian estimation in Generalized additive models (GAMs) by unifiying P-splines and Laplace approximations. 
The Laplace-P-spline model is endowed with closed form expressions for the gradient and Hessian of the log posterior penalty vector. These analytical forms constitute a valuable asset for a computationally efficient and precise exploration strategy of the posterior penalty space that in turn leads to an accurate approximation of the joint posterior latent field (including the regression and spline parameters in the generalized additive model) even when the number of smooth functions is large.\\
\indent Extensive simulation studies show that the algorithms underlying LPS and LPSMAP exhibit good estimation quality with respect to the considered performance metrics, as shown for instance by non-significant biases or frequentist coverage probability of credible intervals appreciably close to their nominal value. Furthermore, our approximate Bayesian approach has proved to be reliable in terms of estimation performance with respect to smooth additive terms.\\
\indent Finally, even though the Laplace-P-spline approach works from a complete Bayesian perspective, the computational budget required for inference is relatively low as compared to existing methods fully relying on MCMC algorithms. A future research challenge will be to summarize the algorithms in a software package to disseminate the LPS and LPSMAP approaches. Moreover, it would be interesting to explore the idea to handle models for spatial data or with additional hierarchy levels.

\vspace{0.3cm}

\subsection*{Acknowledgments}

Gressani Oswaldo wants to thank the Luxembourgish Ministry of higher education and research for a PhD program grant. The authors are also grateful to Dr. Th\'er\`ese Stukel for granting permission to use the nutritional study data in this article.

\vspace{0.5cm}

\section{Appendices}

\subsection{Appendix A1}

\vspace{-0.2cm}

\noindent This appendix provides in full details the analytical derivations of the gradient and Hessian associated to the (log-) posterior of the log penalty vector:

\vspace{-0.3cm}

\begin{eqnarray} \label{target}
\log \tilde{p}(\mathbf{v} \vert \mathcal{D}) &\dot{=}& -\frac{1}{2} \underbrace{\log \vert B^T \widetilde{W} B+Q_{\boldsymbol{\xi}}^{\mathbf{v}} \vert}_{\text{Term I}}+\underbrace{\Bigg(\frac{\nu+K-1}{2}\Bigg)\sum_{j=1}^q v_j}_{\text{Term II}}+\underbrace{\frac{1}{\varkappa} \sum_{i=1}^n y_i \mathbf{b}_i^T \mtilde \varpitilde}_{\text{Term III}} \nonumber \\
&&-\underbrace{\frac{1}{\varkappa} \sum_{i=1}^n s\left(\mathbf{b}_i^T \mtilde \varpitilde \right)}_{\text{Term IV}}-\frac{1}{2} \underbrace{\varpitilde^T \mtilde Q_{\boldsymbol{\xi}}^{\mathbf{v}} \mtilde \varpitilde}_{\text{Term V}} \nonumber \\
&&-\Big(\frac{ \nu}{2}+a_{\delta}\Big)\underbrace{\sum_{j=1}^q \log\Big(b_{\delta}+\frac{\nu}{2} \exp(v_j) \Big)}_{\text{Term VI}},
\end{eqnarray}

\vspace{0.3cm}

\noindent where for notational convenience, we define $\widetilde{\mathcal{M}}_{\boldsymbol{\xi}}^{\mathbf{v}}:=\left(B^T \widetilde{W} B+ Q_{\boldsymbol{\xi}}^{\mathbf{v}}\right)^{-1}$.

\subsection*{Gradient associated to the penalty in a GAM}

\noindent To obtain the gradient of $\log \tilde{p}(\mathbf{v} \vert \mathcal{D})$, the partial derivatives of the latter quantity with respect to $v_j, \ j=1,\dots,q$ are required. The partial derivative of Term I in (\ref{target}) can be obtained using Jacobi's formula:

\vspace{-0.3cm}

\begin{eqnarray} \label{TermI}
\frac{\partial \log \vert B^T \widetilde{W} B+Q^{\mathbf{v}}_{\boldsymbol{\xi}} \vert}{\partial v_j}&=&\frac{1}{\vert B^T\widetilde{W}B+Q^{\mathbf{v}}_{\boldsymbol{\xi}} \vert} \frac{\partial}{\partial v_j} \vert B^T\widetilde{W}B+Q^{\mathbf{v}}_{\boldsymbol{\xi}} \vert \nonumber \\
&=& \frac{1}{\vert B^T\widetilde{W}B+Q^{\mathbf{v}}_{\boldsymbol{\xi}} \vert} \text{Tr}\left( \text{adj}(B^T\widetilde{W}B+Q^{\mathbf{v}}_{\boldsymbol{\xi}}) \frac{\partial}{\partial v_j} (B^T\widetilde{W}B+Q^{\mathbf{v}}_{\boldsymbol{\xi}}) \right) \nonumber \\
&=&\frac{1}{\vert B^T\widetilde{W}B+Q^{\mathbf{v}}_{\boldsymbol{\xi}} \vert} \text{Tr}\Big( \vert B^T\widetilde{W}B+Q^{\mathbf{v}}_{\boldsymbol{\xi}} \vert\  (B^T\widetilde{W}B+Q^{\mathbf{v}}_{\boldsymbol{\xi}})^{-1} \nonumber \\
&& \hspace{3.2cm} \frac{\partial}{\partial v_j} (B^T\widetilde{W}B+Q^{\mathbf{v}}_{\boldsymbol{\xi}}) \Big) \nonumber \\
&=& \text{Tr}\left(\mtilde \widetilde{P}_{v_j} \right), \nonumber
\end{eqnarray}

\vspace{-0.1cm}

\noindent where $\widetilde{P}_{v_j}$ is a (symmetric) block diagonal matrix defined as:

\vspace{-0.4cm}

\[\widetilde{P}_{v_j}:=\frac{\partial}{\partial v_j} (B^T\widetilde{W}B+Q^{\mathbf{v}}_{\boldsymbol{\xi}})=\begin{pmatrix}
0_{p+1,p+1} & 0_{p+1,q\times(K-1)} \\
0_{q\times(K-1),p+1} & \text{diag}(0,\dots,\exp(v_j),\dots,0) \otimes P
\end{pmatrix}.\]

\vspace{0.2cm}

\noindent Derivation of Term II with respect to $v_j$ simply equals the scalar $(\nu+K-1)/2$:

\vspace{-0.2cm}

\begin{eqnarray} \label{Term_II}
\frac{\partial }{\partial v_j} \Bigg(\frac{\nu+K-1}{2}\Bigg) \sum_{j=1}^q v_j=\Bigg(\frac{\nu+K-1}{2}\Bigg). \nonumber 
\end{eqnarray}

\vspace{0.2cm}

\noindent Partial derivatives of Term III and Term IV are obtained using the following result:

\vspace{-0.3cm}

\begin{eqnarray}
\frac{\partial}{\partial v_j} \mtilde&=& \frac{\partial}{\partial v_j} \left(B^T \widetilde{W}B+Q_{\boldsymbol{\xi}}^{\mathbf{v}}\right)^{-1} \nonumber \\
&=& -\left(B^T \widetilde{W}B+Q_{\boldsymbol{\xi}}^{\mathbf{v}}\right)^{-1} \widetilde{P}_{v_j} \left(B^T \widetilde{W}B+Q_{\boldsymbol{\xi}}^{\mathbf{v}}\right)^{-1} \nonumber \\
&=&-\mtilde \widetilde{P}_{v_j} \mtilde. \nonumber
\end{eqnarray}

\vspace{0.2cm}

\noindent Hence for Term III, using the property that the trace is invariant under cyclic permutations:

\vspace{-0.3cm}

\begin{eqnarray} \label{gam_Term_III}
\frac{\partial}{\partial v_j} \left(\frac{1}{\varkappa} \sum_{i=1}^n y_i \mathbf{b}_i^T \mtilde \varpitilde\right)&=&\frac{\partial}{\partial v_j} \text{Tr} \left(\frac{1}{\varkappa} \sum_{i=1}^n y_i \mathbf{b}_i^T \mtilde \varpitilde \right) \nonumber \\
&=& \frac{\partial}{\partial v_j} \left( \frac{1}{\varkappa} \sum_{i=1}^n y_i \text{Tr}\left(\mathbf{b}_i^T \mtilde \varpitilde\right) \right) \nonumber 
\end{eqnarray}

\newpage 

\phantom{k}
\vspace{-1.5cm}

\begin{eqnarray}
&=& \frac{\partial}{\partial v_j} \left( \frac{1}{\varkappa} \sum_{i=1}^n y_i \text{Tr}\left(\varpitilde \mathbf{b}_i^T \mtilde \right) \right) \nonumber \\
&=& \frac{1}{\varkappa} \sum_{i=1}^n y_i \frac{\partial}{\partial v_j} \text{Tr}\left(\varpitilde \mathbf{b}_i^T  \mtilde \right) \nonumber \\
&=& \frac{1}{\varkappa} \sum_{i=1}^n y_i \text{Tr}\left(\varpitilde \mathbf{b}_i^T \frac{\partial}{\partial v_j} \mtilde \right) \nonumber \\
&=&-\frac{1}{\varkappa} \sum_{i=1}^n y_i \text{Tr}\left( \varpitilde \mathbf{b}_i^T \mtilde \widetilde{P}_{v_j} \mtilde \right) \nonumber \\
&=&-\frac{1}{\varkappa} \sum_{i=1}^n y_i \text{Tr}\left(\mathbf{b}_i^T \mtilde \widetilde{P}_{v_j} \mtilde \varpitilde\right) \nonumber \\
&=&-\frac{1}{\varkappa} \sum_{i=1}^n y_i \mathbf{b}_i^T \mtilde \widetilde{P}_{v_j} \mtilde \varpitilde. 
\end{eqnarray}

\noindent For Term IV we use the chain rule and obtain:

\vspace{-0.2cm}

\begin{eqnarray}
\frac{\partial}{\partial v_j} \left(\frac{1}{\varkappa} \sum_{i=1}^n s\left(\mathbf{b}_i^T \mtilde \varpitilde \right) \right)&=&\frac{1}{\varkappa} \sum_{i=1}^n s'\left(\mathbf{b}_i^T \mtilde \varpitilde \right) \frac{\partial}{\partial v_j} \left(\mathbf{b}_i^T \mtilde \varpitilde \right) \nonumber \\
&=& \frac{1}{\varkappa} \sum_{i=1}^n s'\left(\mathbf{b}_i^T \mtilde \varpitilde \right) \frac{\partial}{\partial v_j} \text{Tr}\left(\mathbf{b}_i^T \mtilde \varpitilde \right) \nonumber \\
&=& \frac{1}{\varkappa} \sum_{i=1}^n s'\left(\mathbf{b}_i^T \mtilde \varpitilde \right) \frac{\partial}{\partial v_j} \text{Tr}\left(\varpitilde \mathbf{b}_i^T \mtilde  \right) \nonumber \\
&=& -\frac{1}{\varkappa} \sum_{i=1}^n s'\left(\mathbf{b}_i^T \mtilde \varpitilde \right)  \mathbf{b}_i^T \mtilde \widetilde{P}_{v_j} \mtilde \varpitilde. \nonumber
\end{eqnarray}

\vspace{0.2cm}

\noindent The partial derivative of Term V is obtained as follows:

\vspace{-0.5cm}

\begin{eqnarray}
\frac{\partial}{\partial v_j} \left( \varpitilde^T \mtilde Q_{\boldsymbol{\xi}}^{\mathbf{v}} \mtilde \varpitilde \right)&=&\frac{\partial}{\partial v_j} \text{Tr}\left( \varpitilde^T \mtilde Q_{\boldsymbol{\xi}}^{\mathbf{v}} \mtilde \varpitilde \right) \nonumber \\
&=& \frac{\partial}{\partial v_j} \text{Tr}\left( \varpitilde \varpitilde^T \mtilde Q_{\boldsymbol{\xi}}^{\mathbf{v}} \mtilde \right) \nonumber \\
&=& \text{Tr}\left( \varpitilde \varpitilde^T \frac{\partial}{\partial v_j} \left( \mtilde Q_{\boldsymbol{\xi}}^{\mathbf{v}} \mtilde \right) \right) \nonumber \\
&=& \text{Tr}\left( \varpitilde \varpitilde^T \left(
\frac{\partial \mtilde}{\partial v_j} Q_{\boldsymbol{\xi}}^{\mathbf{v}} \mtilde+\mtilde \frac{\partial Q_{\boldsymbol{\xi}}^{\mathbf{v}}}{\partial v_j} \mtilde +\mtilde Q_{\boldsymbol{\xi}}^{\mathbf{v}} \frac{\partial \mtilde}{\partial v_j} \right) \right) \nonumber 
\end{eqnarray}

\begin{eqnarray}
&=& \text{Tr}\left( \varpitilde \varpitilde^T \left(
-\mtilde \widetilde{P}_{v_j} \mtilde Q_{\boldsymbol{\xi}}^{\mathbf{v}} \mtilde+\mtilde \widetilde{P}_{v_j} \mtilde-\mtilde Q_{\boldsymbol{\xi}}^{\mathbf{v}} \mtilde \widetilde{P}_{v_j} \mtilde \right) \right) \nonumber \\
&=& \text{Tr}\left(\varpitilde^T \left(
-\mtilde \widetilde{P}_{v_j} \mtilde Q_{\boldsymbol{\xi}}^{\mathbf{v}} \mtilde+\mtilde \widetilde{P}_{v_j} \mtilde-\mtilde Q_{\boldsymbol{\xi}}^{\mathbf{v}} \mtilde \widetilde{P}_{v_j} \mtilde \right) \varpitilde \right) \nonumber \\
&=& - \varpitilde^T \mtilde \widetilde{P}_{v_j} \mtilde Q_{\boldsymbol{\xi}}^{\mathbf{v}} \mtilde \varpitilde-\varpitilde^T \mtilde Q_{\boldsymbol{\xi}}^{\mathbf{v}} \mtilde \widetilde{P}_{v_j} \mtilde \varpitilde+ \varpitilde^T \mtilde \widetilde{P}_{v_j} \mtilde \varpitilde \nonumber \\
&=& - \varpitilde^T \mtilde \widetilde{P}_{v_j} \mtilde Q_{\boldsymbol{\xi}}^{\mathbf{v}} \mtilde \varpitilde-\left(\varpitilde^T \mtilde Q_{\boldsymbol{\xi}}^{\mathbf{v}} \mtilde \widetilde{P}_{v_j} \mtilde \varpitilde\right)^T \nonumber \\
&&+\varpitilde^T \mtilde \widetilde{P}_{v_j} \mtilde \varpitilde \nonumber \\
&=& - \varpitilde^T \mtilde \widetilde{P}_{v_j} \mtilde Q_{\boldsymbol{\xi}}^{\mathbf{v}} \mtilde \varpitilde-\varpitilde^T \mtilde \widetilde{P}_{v_j} \mtilde Q_{\boldsymbol{\xi}}^{\mathbf{v}} \mtilde \varpitilde \nonumber \\
&&+ \varpitilde^T \mtilde \widetilde{P}_{v_j} \mtilde \varpitilde \nonumber \\
&=& - 2 \varpitilde^T \mtilde \widetilde{P}_{v_j} \mtilde Q_{\boldsymbol{\xi}}^{\mathbf{v}} \mtilde \varpitilde+\varpitilde^T \mtilde \widetilde{P}_{v_j} \mtilde \varpitilde. \nonumber
\end{eqnarray}

\vspace{0.3cm}

With regard to the derivative of Term VI we have:

\vspace{-0.2cm}

\begin{eqnarray} \label{Term_IV}
\frac{\partial}{\partial v_j} \sum_{j=1}^q\log\Big(b_{\delta}+\frac{\nu}{2} \exp(v_j) \Big)&=& \frac{\frac{\nu}{2} \exp(v_j)}{b_{\delta}+\frac{\nu}{2} \exp(v_j)} \nonumber \\
&=&\frac{1}{1+\frac{2 b_{\delta}}{\nu \exp(v_j)}}. \nonumber 
\end{eqnarray}

\vspace{0.2cm}

\noindent For notational convenience we define $\widetilde{\Upsilon}^{j}_{\mathbf{v}}:=\mtilde \widetilde{P}_{v_j} \mtilde$. From all the above intermediate results for Terms I-VI, the gradient $\nabla_{\mathbf{v}} \log \tilde{p}(\mathbf{v} \vert \mathcal{D})$ has the following entries:

\vspace{-0.4cm}

\newcommand{\tildeups}{\widetilde{\Upsilon}^{j}_{\mathbf{v}}}

\begin{eqnarray}
\frac{\partial \log \tilde{p}(\mathbf{v} \vert \mathcal{D})}{\partial v_j}&=&-\frac{1}{2} \underbrace{\text{Tr}\left(\mtilde \widetilde{P}_{v_j} \right)}_{\text{Term VII}}+\left(\frac{\nu+K-1}{2} \right)-\underbrace{\frac{1}{\varkappa} \sum_{i=1}^n y_i \mathbf{b}_i^T \tildeups \varpitilde}_{\text{Term VIII}} \nonumber \\
&&+\underbrace{\frac{1}{\varkappa} \sum_{i=1}^n s'\left(\mathbf{b}_i^T \mtilde \varpitilde \right)  \mathbf{b}_i^T \tildeups \varpitilde}_{\text{Term IX}}+\underbrace{\varpitilde^T \tildeups Q_{\boldsymbol{\xi}}^{\mathbf{v}} \mtilde \varpitilde}_{\text{Term X}} \nonumber \\
&&-\frac{1}{2} \underbrace{\varpitilde^T \tildeups \varpitilde}_{\text{Term XI}}-\underbrace{\frac{\big(\frac{\nu}{2}+a_{\delta}\big)}{1+\frac{2 b_{\delta}}{\nu \exp(v_j)}}}_{\text{Term XII}}, \ j=1,\dots,q. \nonumber
\end{eqnarray}

\newpage 

\subsection*{Hessian associated to the penalty in a GAM}

\subsubsection*{Diagonal elements}

\vspace{0.2cm}

\noindent First, we focus on the diagonal entries. The derivative of Term VII is:

\vspace{-0.2cm}

\begin{eqnarray} \label{partial_trace}
\frac{\partial}{\partial v_j}  \text{Tr}\left( (B^T\widetilde{W}B+Q^{\mathbf{v}}_{\boldsymbol{\xi}})^{-1} \widetilde{P}_{v_j} \right)&=&\text{Tr}\left(\frac{\partial}{\partial v_j} (B^T\widetilde{W}B+Q^{\mathbf{v}}_{\boldsymbol{\xi}})^{-1} \widetilde{P}_{v_j} \right) \nonumber \\
&=& \text{Tr}\left(-\mtilde \widetilde{P}_{v_j} \mtilde \widetilde{P}_{v_j}+\mtilde \widetilde{P}_{v_j} \right) \nonumber \\
&=& -\text{Tr}\left( \Big(\mtilde \widetilde{P}_{v_j} \Big)^2-\mtilde\widetilde{P}_{v_j} \right). \nonumber 
\end{eqnarray}

\vspace{0.2cm}

\noindent Let us derive the intermediate result:

\vspace{-0.3cm}

\begin{eqnarray} \label{gam_intermediate_res}
\frac{\partial \tildeups}{\partial v_j} &=& \frac{\partial}{\partial v_j} \mtilde \widetilde{P}_{v_j} \mtilde \nonumber \\
&=&\left(
\frac{\partial \mtilde}{\partial v_j} \widetilde{P}_{v_j} \mtilde+\mtilde \frac{\partial \widetilde{P}_{v_j}}{\partial v_j} \mtilde +\mtilde \widetilde{P}_{v_j} \frac{\partial \mtilde}{\partial v_j} \right) \nonumber \\
&=& \left(-\mtilde \widetilde{P}_{v_j} \mtilde \widetilde{P}_{v_j} \mtilde+ \mtilde \widetilde{P}_{v_j} \mtilde- \mtilde \widetilde{P}_{v_j} \mtilde \widetilde{P}_{v_j} \mtilde \right) \nonumber \\
&=& \left(-2 \left( \mtilde \widetilde{P}_{v_j} \right)^2 \mtilde+\tildeups \right). 
\end{eqnarray}

\noindent Partial differentiation of Term VIII yields:

\vspace{-0.2cm}

\begin{eqnarray}
\frac{\partial}{\partial v_j} \left(\frac{1}{\varkappa} \sum_{i=1}^n y_i \mathbf{b}_i^T \tildeups \varpitilde \right)&=&\frac{\partial}{\partial v_j} \text{Tr}\left(\frac{1}{\varkappa} \sum_{i=1}^n y_i \mathbf{b}_i^T \tildeups \varpitilde \right) \nonumber \\
&=& \frac{\partial}{\partial v_j} \left(\frac{1}{\varkappa} \sum_{i=1}^n y_i \text{Tr}\left(\mathbf{b}_i^T \tildeups \varpitilde \right) \right) \nonumber \\
&=& \frac{\partial}{\partial v_j} \left(\frac{1}{\varkappa} \sum_{i=1}^n y_i \text{Tr}\left(\varpitilde \mathbf{b}_i^T \tildeups \right) \right) \nonumber \\
&=&  \frac{1}{\varkappa} \sum_{i=1}^n y_i \frac{\partial}{\partial v_j} \text{Tr}\left(\varpitilde \mathbf{b}_i^T \tildeups \right)  \nonumber \\
&=&  \frac{1}{\varkappa} \sum_{i=1}^n y_i  \text{Tr}\left(\varpitilde \mathbf{b}_i^T \left(\frac{\partial \tildeups }{\partial v_j} \right) \right), \nonumber
\end{eqnarray}

\newpage

\noindent and using intermediate result (\ref{gam_intermediate_res}), one obtains for Term VIII:

\vspace{-0.1cm}

\begin{eqnarray}
\frac{\partial}{\partial v_j} \left(\frac{1}{\varkappa} \sum_{i=1}^n y_i \mathbf{b}_i^T \tildeups \varpitilde \right)&=&-\frac{1}{\varkappa} \sum_{i=1}^n y_i \text{Tr} \left(\varpitilde \mathbf{b}_i^T \left(2 \left( \mtilde \widetilde{P}_{v_j} \right)^2 \mtilde-\tildeups \right) \right) \nonumber \\
&=& -\frac{1}{\varkappa} \sum_{i=1}^n y_i \text{Tr} \left(\mathbf{b}_i^T \left(2 \left( \mtilde \widetilde{P}_{v_j} \right)^2 \mtilde-\tildeups \right) \varpitilde \right) \nonumber \\
&=& -\frac{1}{\varkappa} \sum_{i=1}^n y_i \mathbf{b}_i^T \left(2 \left( \mtilde \widetilde{P}_{v_j} \right)^2 \mtilde-\tildeups \right) \varpitilde. \nonumber 
\end{eqnarray}

\vspace{0.2cm}

\noindent For Term IX, we have:

\vspace{-0.2cm}

\begin{eqnarray}
\frac{\partial}{\partial v_j} \left(\frac{1}{\varkappa} \sum_{i=1}^n s'\left(\mathbf{b}_i^T \mtilde \varpitilde \right)  \mathbf{b}_i^T \tildeups \varpitilde\right)&=&\frac{1}{\varkappa} \sum_{i=1}^n \Bigg(
s''(\mathbf{b}_i^T \mtilde \varpitilde) \frac{\partial}{\partial v_j} \text{Tr}\left(\mathbf{b}_i^T \mtilde \varpitilde \right) \left(\mathbf{b}_i^T \tildeups \varpitilde \right)\nonumber \\
&&+s'(\mathbf{b}_i^T \mtilde \varpitilde) \frac{\partial}{\partial v_j} \text{Tr}\left(\mathbf{b}_i^T \tildeups \varpitilde \right) \Bigg). \nonumber 
\end{eqnarray}

\noindent Using (\ref{gam_Term_III}) and intermediate result (\ref{gam_intermediate_res}) we have for Term IX:

\vspace{-0.2cm}

\begin{eqnarray}
\frac{\partial}{\partial v_j} \left(\frac{1}{\varkappa} \sum_{i=1}^n s'\left(\mathbf{b}_i^T \mtilde \varpitilde \right)  \mathbf{b}_i^T \tildeups \varpitilde\right) \hspace{-0.2cm} &=& \hspace{-0.2cm} \frac{1}{\varkappa} \sum_{i=1}^n \Bigg(
s''(\mathbf{b}_i^T \mtilde \varpitilde)  \left(-\mathbf{b}_i^T \tildeups \varpitilde \right) \left(\mathbf{b}_i^T \tildeups \varpitilde \right) \nonumber \\
&&+s'(\mathbf{b}_i^T \mtilde \varpitilde)\mathbf{b}_i^T \left(-2 \left( \mtilde \widetilde{P}_{v_j} \right)^2 \mtilde+\tildeups \right) \varpitilde \Bigg) \nonumber \\
&=&  \hspace{-0.2cm} -\frac{1}{\varkappa} \sum_{i=1}^n \Bigg(
s'(\mathbf{b}_i^T \mtilde \varpitilde)\mathbf{b}_i^T \left(2 \left( \mtilde \widetilde{P}_{v_j} \right)^2 \mtilde-\tildeups \right) \varpitilde \nonumber \\
&&+s''(\mathbf{b}_i^T \mtilde \varpitilde) \left(\mathbf{b}_i^T \tildeups \varpitilde \right)^2 \Bigg). \nonumber
\end{eqnarray}

\noindent The partial derivative of Term X is obtained as follows:

\vspace{-0.3cm}

\begin{eqnarray}
\frac{\partial}{\partial v_j} \left(\varpitilde^T \tildeups Q_{\boldsymbol{\xi}}^{\mathbf{v}} \mtilde \varpitilde \right)&=&\frac{\partial}{\partial v_j} \text{Tr}\left(\varpitilde^T \tildeups Q_{\boldsymbol{\xi}}^{\mathbf{v}} \mtilde \varpitilde \right) \nonumber \\
&=&\frac{\partial}{\partial v_j} \text{Tr}\left(\varpitilde \varpitilde^T \tildeups Q_{\boldsymbol{\xi}}^{\mathbf{v}} \mtilde  \right) \nonumber \\
&=& \text{Tr}\left( \varpitilde \varpitilde^T \frac{\partial}{\partial v_j} \left(\tildeups Q_{\boldsymbol{\xi}}^{\mathbf{v}} \mtilde  \right) \right) \nonumber 
\end{eqnarray}

\begin{eqnarray}
&=& \text{Tr}\Bigg( \varpitilde \varpitilde^T \Bigg( \frac{\partial \tildeups}{\partial v_j} Q_{\boldsymbol{\xi}}^{\mathbf{v}} \mtilde+ \tildeups \frac{\partial Q_{\boldsymbol{\xi}}^{\mathbf{v}}}{\partial v_j} \mtilde+\tildeups Q_{\boldsymbol{\xi}}^{\mathbf{v}} \frac{\partial \mtilde}{\partial v_j} \Bigg) \Bigg) \nonumber \\
&=& \text{Tr}\Bigg( \varpitilde \varpitilde^T \Bigg(
\Big( -2 \left(\mtilde \widetilde{P}_{v_j} \right)^2 \mtilde +\tildeups \Big)  Q_{\boldsymbol{\xi}}^{\mathbf{v}} \mtilde \nonumber \\
&&+ \tildeups \widetilde{P}_{v_j}  \mtilde- \tildeups Q_{\boldsymbol{\xi}}^{\mathbf{v}} \tildeups \Bigg) \Bigg) \nonumber \\
&=& \text{Tr}\Bigg( \varpitilde^T \Bigg(
-2 \left(\mtilde \widetilde{P}_{v_j} \right)^2 \mtilde Q_{\boldsymbol{\xi}}^{\mathbf{v}} \mtilde+\tildeups Q_{\boldsymbol{\xi}}^{\mathbf{v}} \mtilde \nonumber \\
&&+ \tildeups \widetilde{P}_{v_j} \mtilde- \tildeups Q_{\boldsymbol{\xi}}^{\mathbf{v}} \tildeups \Bigg) \varpitilde \Bigg) \nonumber \\
&=& -2 \varpitilde^T \left(\mtilde \widetilde{P}_{v_j} \right)^2 \mtilde Q_{\boldsymbol{\xi}}^{\mathbf{v}} \mtilde \varpitilde+ \varpitilde^T \tildeups \left(Q_{\boldsymbol{\xi}}^{\mathbf{v}}+\widetilde{P}_{v_j}\right) \mtilde \varpitilde \nonumber \\
&&-\varpitilde^T \tildeups Q_{\boldsymbol{\xi}}^{\mathbf{v}} \tildeups \varpitilde. \nonumber 
\end{eqnarray}

\noindent Partial differentiation of Term XI gives us:

\vspace{-0.3cm}

\begin{eqnarray}
\frac{\partial}{\partial v_j} \left(\varpitilde^T \tildeups \varpitilde \right)&=&\frac{\partial}{\partial v_j} \text{Tr}\left(\varpitilde^T \tildeups \varpitilde \right) \nonumber \\
&=&\frac{\partial}{\partial v_j} \text{Tr}\left(\varpitilde \varpitilde^T \tildeups \right) \nonumber \\
&=& \text{Tr}\left(\varpitilde \varpitilde^T \frac{\partial \tildeups}{\partial v_j} \right) \nonumber \\
&=& \text{Tr}\left(\varpitilde \varpitilde^T \left(-2 \left(\mtilde \widetilde{P}_{v_j} \right)^2 \mtilde+\tildeups \right) \right) \nonumber \\
&=& \text{Tr}\left( \varpitilde^T \left(-2 \left(\mtilde \widetilde{P}_{v_j} \right)^2 \mtilde+\tildeups \right) \varpitilde \right) \nonumber \\
&=&-2 \varpitilde^T \left(\mtilde \widetilde{P}_{v_j} \right)^2 \mtilde \varpitilde+ \varpitilde^T \tildeups \varpitilde. \nonumber
\end{eqnarray}

\noindent Finally derivation of Term XII gives us:

\vspace{-0.5cm}

\begin{eqnarray} 
\frac{\partial}{\partial v_j} \frac{\big(\frac{\nu}{2}+a_{\delta}\big)}{\Big(1+\frac{2 b_{\delta}}{\nu \exp(v_j)}\Big)}=\frac{b_{\delta} \big(1+\frac{2 a_{\delta}}{\nu}\big) \exp(-v_j)}{\Big(1+\frac{2 b_{\delta}}{\nu \exp(v_j)}\Big)^2}. \nonumber
\end{eqnarray}

\newpage 

\noindent Using the differentiation results for Terms VII-XII, the diagonal elements of the Hessian of $\log \tilde{p}(\mathbf{v} \vert \mathcal{D})$ are:

\begin{eqnarray}
\frac{\partial^2 \log \tilde{p}(\mathbf{v} \vert \mathcal{D})}{\partial v_j^2}&=&\frac{1}{2} \text{Tr}\left( \left(\mtilde \widetilde{P}_{v_j} \right)^2-\mtilde \widetilde{P}_{v_j} \right)+\frac{1}{\varkappa} \sum_{i=1}^n y_i \mathbf{b}_i^T \left(2 \left( \mtilde \widetilde{P}_{v_j} \right)^2 \mtilde-\tildeups \right) \varpitilde \nonumber \\
&& \hspace{-1.8cm} -\frac{1}{\varkappa} \sum_{i=1}^n \Bigg(
s'(\mathbf{b}_i^T \mtilde \varpitilde)\mathbf{b}_i^T \left(2 \left( \mtilde \widetilde{P}_{v_j} \right)^2 \mtilde-\tildeups \right) \varpitilde+s''(\mathbf{b}_i^T \mtilde \varpitilde) \left(\mathbf{b}_i^T \tildeups \varpitilde \right)^2 \Bigg) \nonumber \\
&& \hspace{-1.8cm} -2 \varpitilde^T \left(\mtilde \widetilde{P}_{v_j} \right)^2 \mtilde Q_{\boldsymbol{\xi}}^{\mathbf{v}} \mtilde \varpitilde+ \varpitilde^T \tildeups \left(Q_{\boldsymbol{\xi}}^{\mathbf{v}}+\widetilde{P}_{v_j}\right) \mtilde \varpitilde- \varpitilde^T \tildeups Q_{\boldsymbol{\xi}}^{\mathbf{v}} \tildeups \varpitilde \nonumber \\
&&  \hspace{-1.8cm} +\varpitilde^T \left(\mtilde \widetilde{P}_{v_j} \right)^2 \mtilde \varpitilde- \frac{1}{2} \varpitilde^T \tildeups \varpitilde-\frac{b_{\delta} \big(1+\frac{2 a_{\delta}}{\nu}\big) \exp(-v_j)}{\Big(1+\frac{2 b_{\delta}}{\nu \exp(v_j)}\Big)^2}, \ \ j=1,\dots,q. \nonumber
\end{eqnarray}

\subsubsection*{Off-diagonal elements}

\noindent Note that for index $s\neq j$ we have for Term VII:

\vspace{-0.4cm}

\begin{eqnarray}
\frac{\partial}{\partial v_s} \text{Tr}\left(\mtilde \widetilde{P}_{v_j} \right)&=&\text{Tr}\left( \frac{\partial \mtilde}{\partial v_s} \widetilde{P}_{v_j} \right) \nonumber \\
&=&-\text{Tr}\left(\mtilde \widetilde{P}_{v_s} \mtilde \widetilde{P}_{v_j} \right). \nonumber
\end{eqnarray}

\vspace{0.1cm}

\newcommand{\tildeupss}{\widetilde{\Upsilon}^{s}_{\mathbf{v}}}

\noindent Let us define $\widetilde{\Upsilon}^{s}_{\mathbf{v}}:=\mtilde \widetilde{P}_{v_s} \mtilde$ and consider the following intermediate result:

\vspace{-0.2cm}

\begin{eqnarray} \label{gam_intermediate_res2}
\frac{\partial \tildeups}{\partial v_s} &=& \frac{\partial}{\partial v_s} \mtilde \widetilde{P}_{v_j} \mtilde \nonumber \\
&=&\left(
\frac{\partial \mtilde}{\partial v_s} \widetilde{P}_{v_j} \mtilde+\mtilde \frac{\partial \widetilde{P}_{v_j}}{\partial v_s} \mtilde +\mtilde \widetilde{P}_{v_j} \frac{\partial \mtilde}{\partial v_s} \right) \nonumber \\
&=& \left(-\mtilde \widetilde{P}_{v_s} \mtilde \widetilde{P}_{v_j} \mtilde- \mtilde \widetilde{P}_{v_j} \mtilde \widetilde{P}_{v_s} \mtilde \right) \nonumber \\
&=& -\left(\tildeupss \widetilde{P}_{v_j} \mtilde+\mtilde \widetilde{P}_{v_j} \tildeupss \right).  
\end{eqnarray}

\vspace{0.2cm}

\noindent Result (\ref{gam_intermediate_res2}) can be used to obtain the differentiation of Term VIII:

\vspace{-0.6cm}

\begin{eqnarray}
\frac{\partial}{\partial v_s} \left( \frac{1}{\varkappa} \sum_{i=1}^n y_i \mathbf{b}_i^T \tildeups \varpitilde \right)&=&\frac{\partial}{\partial v_s} \text{Tr}\left( \frac{1}{\varkappa} \sum_{i=1}^n y_i \mathbf{b}_i^T \tildeups \varpitilde \right) \nonumber \\
&=& \frac{\partial}{\partial v_s} \left( \frac{1}{\varkappa} \sum_{i=1}^n y_i \text{Tr}\left(\mathbf{b}_i^T \tildeups \varpitilde \right) \right) \nonumber \\
&=& \frac{1}{\varkappa} \sum_{i=1}^n y_i  \frac{\partial}{\partial v_s} \text{Tr}\left(\varpitilde \mathbf{b}_i^T \tildeups \right)  \nonumber 
\end{eqnarray}

\begin{eqnarray}
&=& \frac{1}{\varkappa} \sum_{i=1}^n y_i   \text{Tr}\Bigg(\varpitilde \mathbf{b}_i^T \frac{\partial \tildeups}{\partial v_s} \Bigg)  \nonumber \\ 
&=& -\frac{1}{\varkappa} \sum_{i=1}^n y_i   \text{Tr}\Big( \varpitilde \mathbf{b}_i^T \left(\tildeupss \widetilde{P}_{v_j} \mtilde+\mtilde \widetilde{P}_{v_j} \tildeupss \right) \Big) \nonumber \\
&=& -\frac{1}{\varkappa} \sum_{i=1}^n y_i   \text{Tr}\Big(  \mathbf{b}_i^T \left(\tildeupss \widetilde{P}_{v_j} \mtilde+\mtilde \widetilde{P}_{v_j} \tildeupss \right) \varpitilde \Big) \nonumber \\
&=& -\frac{1}{\varkappa} \sum_{i=1}^n y_i  \mathbf{b}_i^T \left(\tildeupss \widetilde{P}_{v_j} \mtilde+\mtilde \widetilde{P}_{v_j} \tildeupss \right) \varpitilde. \nonumber 
\end{eqnarray}

\noindent To derive Term IX, we also use result (\ref{gam_intermediate_res2}):

\vspace{-0.2cm}

\begin{eqnarray}
\frac{\partial}{\partial v_s} \left(\frac{1}{\varkappa} \sum_{i=1}^n s'\left(\mathbf{b}_i^T \mtilde \varpitilde \right)  \mathbf{b}_i^T \tildeups \varpitilde\right) \hspace{-0.3cm} &=& \hspace{-0.3cm} \frac{1}{\varkappa} \sum_{i=1}^n \Bigg(
s''(\mathbf{b}_i^T \mtilde \varpitilde) \frac{\partial}{\partial v_s} \text{Tr}\left(\mathbf{b}_i^T \mtilde \varpitilde \right) \left(\mathbf{b}_i^T \tildeups \varpitilde \right)\nonumber \\
&&+s'(\mathbf{b}_i^T \mtilde \varpitilde) \frac{\partial}{\partial v_s} \text{Tr}\left(\mathbf{b}_i^T \tildeups \varpitilde \right) \Bigg) \nonumber \\
&=& \hspace{-0.3cm} \frac{1}{\varkappa} \sum_{i=1}^n \Bigg(
s''(\mathbf{b}_i^T \mtilde \varpitilde)\left(-\mathbf{b}_i^T \tildeupss \varpitilde \right)\left(\mathbf{b}_i^T \tildeups \varpitilde \right) \nonumber \\
&&+s'(\mathbf{b}_i^T \mtilde \varpitilde) \left(-\mathbf{b}_i^T \left(\tildeupss \widetilde{P}_{v_j} \mtilde+\mtilde \widetilde{P}_{v_j} \tildeupss \right) \varpitilde \right) \nonumber \\
&=&\hspace{-0.3cm} -\frac{1}{\varkappa} \sum_{i=1}^n \Bigg(s'(\mathbf{b}_i^T \mtilde \varpitilde) \mathbf{b}_i^T \left(\tildeupss \widetilde{P}_{v_j} \mtilde+\mtilde \widetilde{P}_{v_j} \tildeupss \right) \varpitilde \nonumber \\
&&+s''(\mathbf{b}_i^T \mtilde \varpitilde)\left(\mathbf{b}_i^T \tildeupss \varpitilde \right)\left(\mathbf{b}_i^T \tildeups \varpitilde \right) \Bigg). \nonumber 
\end{eqnarray}

\vspace{0.2cm}

\noindent Partial differentiation of Term X goes as follows:

\vspace{-0.4cm}

\begin{eqnarray}
\frac{\partial}{\partial v_s} \left(\varpitilde^T \tildeups Q_{\boldsymbol{\xi}}^{\mathbf{v}} \mtilde \varpitilde \right)&=&\frac{\partial}{\partial v_s} \text{Tr}\left(\varpitilde^T \tildeups Q_{\boldsymbol{\xi}}^{\mathbf{v}} \mtilde \varpitilde \right) \nonumber \\
&=&\frac{\partial}{\partial v_s} \text{Tr}\left(\varpitilde \varpitilde^T \tildeups Q_{\boldsymbol{\xi}}^{\mathbf{v}} \mtilde  \right) \nonumber \\
&=& \text{Tr}\left( \varpitilde \varpitilde^T \frac{\partial}{\partial v_s} \left(\tildeups Q_{\boldsymbol{\xi}}^{\mathbf{v}} \mtilde  \right) \right) \nonumber \\
&=& \text{Tr}\Bigg( \varpitilde \varpitilde^T \Bigg( \frac{\partial \tildeups}{\partial v_s} Q_{\boldsymbol{\xi}}^{\mathbf{v}} \mtilde+ \tildeups \frac{\partial Q_{\boldsymbol{\xi}}^{\mathbf{v}}}{\partial v_s} \mtilde+\tildeups Q_{\boldsymbol{\xi}}^{\mathbf{v}} \frac{\partial \mtilde}{\partial v_s} \Bigg) \Bigg) \nonumber 
\end{eqnarray}

\begin{eqnarray}
&=& \text{Tr}\Bigg( \varpitilde \varpitilde^T \Bigg( -\left(\tildeupss \widetilde{P}_{v_j} \mtilde+\mtilde \widetilde{P}_{v_j} \tildeupss \right) Q_{\boldsymbol{\xi}}^{\mathbf{v}} \mtilde \nonumber \\
&&+\tildeups \widetilde{P}_{v_s} \mtilde-\tildeups Q_{\boldsymbol{\xi}}^{\mathbf{v}} \mtilde \widetilde{P}_{v_s} \mtilde \Bigg) \Bigg) \nonumber \\
&=& \text{Tr}\Bigg( \varpitilde^T \Bigg( -\left(\tildeupss \widetilde{P}_{v_j} \mtilde+\mtilde \widetilde{P}_{v_j} \tildeupss \right) Q_{\boldsymbol{\xi}}^{\mathbf{v}} \mtilde \nonumber \\
&&+\tildeups \widetilde{P}_{v_s} \mtilde-\tildeups Q_{\boldsymbol{\xi}}^{\mathbf{v}} \mtilde \widetilde{P}_{v_s} \mtilde \Bigg) \varpitilde  \Bigg) \nonumber \\
&=&- \varpitilde^T \tildeupss \widetilde{P}_{v_j} \mtilde Q_{\boldsymbol{\xi}}^{\mathbf{v}} \mtilde \varpitilde-\varpitilde^T \mtilde \widetilde{P}_{v_j} \tildeupss Q_{\boldsymbol{\xi}}^{\mathbf{v}} \mtilde \varpitilde \nonumber \\
&&+ \varpitilde^T \tildeups \widetilde{P}_{v_s} \mtilde \varpitilde-\varpitilde^T \tildeups Q_{\boldsymbol{\xi}}^{\mathbf{v}}  \tildeupss \varpitilde. \nonumber 
\end{eqnarray}

\noindent Partial differentiation of Term XI gives us:

\vspace{-0.3cm}

\begin{eqnarray}
\frac{\partial}{\partial v_s} \left(\varpitilde^T \tildeups \varpitilde \right)&=&\frac{\partial}{\partial v_s} \text{Tr}\left(\varpitilde^T \tildeups \varpitilde \right) \nonumber \\
&=&\frac{\partial}{\partial v_s} \text{Tr}\left(\varpitilde \varpitilde^T \tildeups \right) \nonumber \\
&=& \text{Tr}\left(\varpitilde \varpitilde^T \frac{\partial \tildeups}{\partial v_s} \right) \nonumber \\
&=& -\text{Tr}\left(\varpitilde \varpitilde^T \left(\tildeupss \widetilde{P}_{v_j} \mtilde+\mtilde \widetilde{P}_{v_j} \tildeupss \right) \right) \nonumber \\
&=& -\text{Tr}\left( \varpitilde^T \left(\tildeupss \widetilde{P}_{v_j} \mtilde+\mtilde \widetilde{P}_{v_j} \tildeupss \right) \varpitilde \right) \nonumber \\ 
&=&- \varpitilde^T \tildeupss \widetilde{P}_{v_j} \mtilde \varpitilde- \left(\varpitilde^T \mtilde \widetilde{P}_{v_j} \tildeupss \varpitilde\right)^T \nonumber \\
&=&- \varpitilde^T \tildeupss \widetilde{P}_{v_j} \mtilde \varpitilde- \varpitilde^T \tildeupss \widetilde{P}_{v_j} \mtilde \varpitilde \nonumber \\
&=&-2\varpitilde^T \tildeupss \widetilde{P}_{v_j} \mtilde \varpitilde. \nonumber 
\end{eqnarray}

\vspace{0.2cm}

\noindent Finally, using the above results, the off-diagonal elements $s=1,\dots,q$; $j=1,\dots,q$ and $s\neq j$ of the Hessian of $\log \tilde{p}(\mathbf{v} \vert \mathcal{D})$ are:

\vspace{-0.2cm}

\begin{eqnarray}
\frac{\partial^2 \log \tilde{p}(\mathbf{v} \vert \mathcal{D})}{\partial v_s\  \partial v_j}&=& \frac{1}{2}\text{Tr}\left(\mtilde \widetilde{P}_{v_s} \mtilde \widetilde{P}_{v_j} \right)+\frac{1}{\varkappa} \sum_{i=1}^n y_i  \mathbf{b}_i^T \left(\tildeupss \widetilde{P}_{v_j} \mtilde+\mtilde \widetilde{P}_{v_j} \tildeupss \right) \varpitilde \nonumber \\
&&-\frac{1}{\varkappa} \sum_{i=1}^n \Bigg(s'(\mathbf{b}_i^T \mtilde \varpitilde) \mathbf{b}_i^T \left(\tildeupss \widetilde{P}_{v_j} \mtilde+\mtilde \widetilde{P}_{v_j} \tildeupss \right) \varpitilde \nonumber 
\end{eqnarray}

\newpage 

\phantom{k}
\vspace{-1.5cm}

\begin{eqnarray}
&&+s''(\mathbf{b}_i^T \mtilde \varpitilde)\left(\mathbf{b}_i^T \tildeupss \varpitilde \right)\left(\mathbf{b}_i^T \tildeups \varpitilde \right) \Bigg) \nonumber \\
&&- \varpitilde^T \tildeupss \widetilde{P}_{v_j} \mtilde Q_{\boldsymbol{\xi}}^{\mathbf{v}} \mtilde \varpitilde-\varpitilde^T \mtilde \widetilde{P}_{v_j} \tildeupss Q_{\boldsymbol{\xi}}^{\mathbf{v}} \mtilde \varpitilde \nonumber \\
&&+ \varpitilde^T \tildeups \widetilde{P}_{v_s} \mtilde \varpitilde-\varpitilde^T \tildeups Q_{\boldsymbol{\xi}}^{\mathbf{v}}  \tildeupss \varpitilde+\varpitilde^T \tildeupss \widetilde{P}_{v_j} \mtilde \varpitilde. \nonumber 
\end{eqnarray}

\vspace{0.2cm}

\noindent To summarize, the gradient and Hessian entries of $\log \tilde{p}(\mathbf{v}\vert \mathcal{D})$ are:\\

\noindent \textbf{Gradient $\nabla_{\mathbf{v}} \log \tilde{p}(\mathbf{v} \vert \mathcal{D})$ entries for} $j=1,\dots,q$:

\vspace{-0.2cm}

\begin{eqnarray}
\frac{\partial \log \tilde{p}(\mathbf{v} \vert \mathcal{D})}{\partial v_j}&=&-\frac{1}{2} \text{Tr}\left(\mtilde \widetilde{P}_{v_j} \right)+\left(\frac{\nu+K-1}{2} \right)-\frac{1}{\varkappa} \sum_{i=1}^n y_i \mathbf{b}_i^T \tildeups \varpitilde\nonumber \\
&&+\frac{1}{\varkappa} \sum_{i=1}^n s'\left(\mathbf{b}_i^T \mtilde \varpitilde \right)  \mathbf{b}_i^T \tildeups \varpitilde+\varpitilde^T \tildeups Q_{\boldsymbol{\xi}}^{\mathbf{v}} \mtilde \varpitilde \nonumber \\
&&-\frac{1}{2} \varpitilde^T \tildeups \varpitilde-\frac{\big(\frac{\nu}{2}+a_{\delta}\big)}{1+\frac{2 b_{\delta}}{\nu \exp(v_j)}}. \nonumber 
\end{eqnarray}

\vspace{0.2cm}

\noindent \textbf{Hessian $\nabla_{\mathbf{v}}^2 \log \tilde{p}(\mathbf{v} \vert \mathcal{D})$, diagonal elements $j=1,\dots,q$}:

\vspace{-0.2cm}

\begin{eqnarray}
\frac{\partial^2 \log \tilde{p}(\mathbf{v} \vert \mathcal{D})}{\partial v_j^2}&=&\frac{1}{2} \text{Tr}\left( \left(\mtilde \widetilde{P}_{v_j} \right)^2-\mtilde \widetilde{P}_{v_j} \right)+\frac{1}{\varkappa} \sum_{i=1}^n y_i \mathbf{b}_i^T \left(2 \left( \mtilde \widetilde{P}_{v_j} \right)^2 \mtilde-\tildeups \right) \varpitilde \nonumber \\
&& \hspace{-1.5cm} -\frac{1}{\varkappa} \sum_{i=1}^n \Bigg(
s'(\mathbf{b}_i^T \mtilde \varpitilde)\mathbf{b}_i^T \left(2 \left( \mtilde \widetilde{P}_{v_j} \right)^2 \mtilde-\tildeups \right) \varpitilde+s''(\mathbf{b}_i^T \mtilde \varpitilde) \left(\mathbf{b}_i^T \tildeups \varpitilde \right)^2 \Bigg) \nonumber \\
&&\hspace{-1.5cm} -2 \varpitilde^T \left(\mtilde \widetilde{P}_{v_j} \right)^2 \mtilde Q_{\boldsymbol{\xi}}^{\mathbf{v}} \mtilde \varpitilde+ \varpitilde^T \tildeups \left(Q_{\boldsymbol{\xi}}^{\mathbf{v}}+\widetilde{P}_{v_j}\right) \mtilde \varpitilde- \varpitilde^T \tildeups Q_{\boldsymbol{\xi}}^{\mathbf{v}} \tildeups \varpitilde \nonumber \\
&&\hspace{-1.5cm} +\varpitilde^T \left(\mtilde \widetilde{P}_{v_j} \right)^2 \mtilde \varpitilde- \frac{1}{2} \varpitilde^T \tildeups \varpitilde-\frac{b_{\delta} \big(1+\frac{2 a_{\delta}}{\nu}\big) \exp(-v_j)}{\Big(1+\frac{2 b_{\delta}}{\nu \exp(v_j)}\Big)^2}. \nonumber 
\end{eqnarray}

\vspace{0.2cm}

\noindent \textbf{Hessian $\nabla_{\mathbf{v}}^2 \log \tilde{p}(\mathbf{v} \vert \mathcal{D})$, off-diagonal elements $s=1,\dots,q; \ j=1,\dots,q $, $j\neq s$}:

\begin{eqnarray}
\frac{\partial^2 \log \tilde{p}(\mathbf{v} \vert \mathcal{D})}{\partial v_s\  \partial v_j}&=& \frac{1}{2}\text{Tr}\left(\mtilde \widetilde{P}_{v_s} \mtilde \widetilde{P}_{v_j} \right)+\frac{1}{\varkappa} \sum_{i=1}^n y_i  \mathbf{b}_i^T \left(\tildeupss \widetilde{P}_{v_j} \mtilde+\mtilde \widetilde{P}_{v_j} \tildeupss \right) \varpitilde \nonumber \\
&&-\frac{1}{\varkappa} \sum_{i=1}^n \Bigg(s'(\mathbf{b}_i^T \mtilde \varpitilde) \mathbf{b}_i^T \left(\tildeupss \widetilde{P}_{v_j} \mtilde+\mtilde \widetilde{P}_{v_j} \tildeupss \right) \varpitilde \nonumber 
\end{eqnarray}

\newpage 

\phantom{k}
\vspace{-1.5cm}

\begin{eqnarray}
&&+s''(\mathbf{b}_i^T \mtilde \varpitilde)\left(\mathbf{b}_i^T \tildeupss \varpitilde \right)\left(\mathbf{b}_i^T \tildeups \varpitilde \right) \Bigg) \nonumber \\
&&- \varpitilde^T \tildeupss \widetilde{P}_{v_j} \mtilde Q_{\boldsymbol{\xi}}^{\mathbf{v}} \mtilde \varpitilde-\varpitilde^T \mtilde \widetilde{P}_{v_j} \tildeupss Q_{\boldsymbol{\xi}}^{\mathbf{v}} \mtilde \varpitilde \nonumber \\
&&+ \varpitilde^T \tildeups \widetilde{P}_{v_s} \mtilde \varpitilde-\varpitilde^T \tildeups Q_{\boldsymbol{\xi}}^{\mathbf{v}}  \tildeupss \varpitilde+\varpitilde^T \tildeupss \widetilde{P}_{v_j} \mtilde \varpitilde. \nonumber 
\end{eqnarray}

\vspace{0.3cm}

\noindent To assess the accuracy of the above gradient and Hessian equations associated to $\log \tilde{p}(\mathbf{v} \vert \mathcal{D})$, we have implemented a procedure in \textbf{R} that compares the analytical results with the numerical derivatives of $\log \tilde{p}(\mathbf{v} \vert \mathcal{D})$ obtained with the {\tt{grad()}} and {\tt{hessian()}} functions of the {\tt{numDeriv}} package at 50 randomly selected points $\mathbf{v} \in \mathbb{R}^3$ with $v_j \sim \mathcal{U}(-4,8), j=1,2,3$ and the response generated from a Poisson distribution. Numerical and analytical derivative results turn out to be very similar, a clear indication that the derived analytical results are accurate.

\subsection{Appendix A2}

\noindent In this appendix, we show the derivations related to the skew-normal fit to the conditional $\tilde{p}(v_j \vert \hat{\mathbf{v}}_{-j},\mathcal{D})$. The skew-normal distribution denoted by $X\sim \text{SN}(\mu, \varsigma^2, \rho)$ has probability density function:

\vspace{-0.8cm}

\begin{eqnarray}
p(x)=\frac{2}{\varsigma} \varphi \left(\frac{x-\mu}{\varsigma} \right)\ \Phi\left(\rho \frac{(x-\mu)}{\varsigma}\right).
\end{eqnarray}

\vspace{0.2cm}

\noindent The first moment and the second and third central moments of $X$ are given by:

\vspace{-0.4cm}

\begin{eqnarray}
E(X)&=&\mu+\varsigma\ \sqrt{\frac{2}{\pi}}\ \psi, \nonumber \\
E\big((X-E(X))^2\big)&=&\varsigma^2 \left(1-\frac{2}{\pi} \ \psi^2\right), \nonumber \\
E\big((X-E(X))^3\big)&=& \frac{1}{2} (4-\pi)\ \varsigma^3\ \left(\frac{2}{\pi}\right)^{\frac{3}{2}}\ \psi^3, \nonumber 
\end{eqnarray}

\vspace{0.2cm}

\noindent where $\psi=\rho/\sqrt{1+\rho^2} \in (-1,1)$. These theoretical moments will be matched with the empirical moments of the the conditional distributions $\tilde{p}(v_j \vert \hat{\mathbf{v}}_{-j},\mathcal{D})$, where $\hat{\mathbf{v}}_{-j}$ is the vector $\hat{\mathbf{v}}$ without the $j^{th}$ entry. The empirical moments of the conditionals are computed on an equidistant grid $\{v_{jl}\}_{l=1}^L$ with interval length $\Delta_l$ and correspond to:

\vspace{-0.4cm}

\begin{eqnarray}
\mathcal{M}_{j1}&=&\sum_{l=1}^L v_{jl}\  \tilde{p}(v_{jl}\vert \hat{\mathbf{v}}_{-j}, \mathcal{D})\ \Delta_l, \nonumber \\
\mathcal{M}_{j2}&=&\sum_{l=1}^L (v_{jl}-\mathcal{M}_{j1})^2\  \tilde{p}(v_{jl} \vert \hat{\mathbf{v}}_{-j}, \mathcal{D})\ \Delta_l, \nonumber \\
\mathcal{M}_{j3}&=&\sum_{l=1}^L (v_{jl}-\mathcal{M}_{j1})^3\  \tilde{p}(v_{jl} \vert  \hat{\mathbf{v}}_{-j}, \mathcal{D})\ \Delta_l. \nonumber \nonumber
\end{eqnarray}

\newpage 

\noindent The skew-normal fit to $\tilde{p}(v_j \vert \hat{\mathbf{v}}_{-j},\mathcal{D})$ is found by matching the empirical and theoretical moments, i.e. the following system needs to be solved:

\vspace{-0.4cm}

\begin{eqnarray}
\mathcal{M}_{j1}&=&\mu+\varsigma\ \sqrt{\frac{2}{\pi}}\ \psi \label{mom1} \\
\mathcal{M}_{j2}&=&\varsigma^2 \left(1-\frac{2}{\pi} \ \psi^2\right)  
\label{mom2} \\
\mathcal{M}_{j3}&=&\frac{1}{2} (4-\pi)\ \varsigma^3\ \Big(\frac{2}{\pi}\Big)^{\frac{3}{2}}\ \psi^3.
\label{mom3}
\end{eqnarray}

\vspace{0.2cm}

\noindent From (\ref{mom2}), we isolate $\varsigma$:

\vspace{-0.4cm}

\begin{eqnarray} \label{scale_isolate}
\varsigma=\sqrt{\frac{\mathcal{M}_{j2}}{\big(1-\frac{2}{\pi} \ \psi^2\big)}}>0.
\end{eqnarray}

\vspace{0.2cm}

\noindent Plugging (\ref{scale_isolate}) in (\ref{mom3}) yields:

\vspace{-0.3cm}

\begin{eqnarray}
\mathcal{M}_{j3}&=&\frac{1}{2} (4-\pi) \frac{\mathcal{M}_{j2}^{\frac{3}{2}}}{\big(1-\frac{2}{\pi} \ \psi^2\big)^{\frac{3}{2}}} \Big(\frac{2}{\pi}\Big)^{\frac{3}{2}} \psi^3 \nonumber \\
&&\Leftrightarrow \frac{\psi^3}{\big(1-\frac{2}{\pi} \ \psi^2\big)^{\frac{3}{2}}}=\frac{2 \mathcal{M}_{j3} \pi^{\frac{3}{2}}}{(4-\pi) \mathcal{M}_{j2}^{\frac{3}{2}} 2^{\frac{3}{2}}} \nonumber \\ 
&&\Leftrightarrow \frac{\psi^3}{\big(1-\frac{2}{\pi} \ \psi^2\big)^{\frac{3}{2}}}= \frac{\mathcal{M}_{j3} \pi^{\frac{3}{2}}}{(4-\pi) \sqrt{2}\  \mathcal{M}_{j2}^{\frac{3}{2}}} \nonumber \\
&&\Leftrightarrow \frac{\psi}{\big(1-\frac{2}{\pi} \ \psi^2\big)^{\frac{1}{2}}}=\frac{\mathcal{M}_{j3}^{\frac{1}{3}} \pi^{\frac{1}{2}}}{(4-\pi)^{\frac{1}{3}} 2^{\frac{1}{6}}\  \mathcal{M}_{j2}^{\frac{1}{2}}}. \nonumber
\end{eqnarray}

\noindent Let $\kappa:=\mathcal{M}_{j3}^{\frac{1}{3}} \pi^{\frac{1}{2}}/(4-\pi)^{\frac{1}{3}} 2^{\frac{1}{6}}\  \mathcal{M}_{j2}^{\frac{1}{2}}$, so that the above equation becomes:

\vspace{-0.4cm}

\begin{eqnarray}
&&\psi=\kappa\ \left(1-\frac{2}{\pi} \psi^2\right)^{\frac{1}{2}} \nonumber \\
&&\Leftrightarrow \psi^2+ \frac{2 \kappa^2}{\pi}\ \psi^2-\kappa^{2}=0 \nonumber \\ 
&&\Leftrightarrow \psi^2 \left(1+\frac{2 \kappa^2}{\pi}\right)-\kappa^2=0. \nonumber
\end{eqnarray}

\noindent The discriminant of the above quadratic equation is $\Delta=4\Big(1+\frac{2 \kappa^2}{\pi}\Big)\kappa^2>0$. Even though there are two solutions, the only solution retained is the one whose sign is the same as the sign of the third empirical central moment. Indeed, if $\mathcal{M}_{j3}$ is negative/positive, $\psi^*$ (and by extension $\rho^*$) should also be negative/positive to capture the negatively/positively skewed pattern of $\tilde{p}(v_j \vert \hat{\mathbf{v}}_{-j},\mathcal{D})$. Hence using the $\text{sign}(\cdot)$ function:

\vspace{-0.3cm}

\begin{eqnarray}
\psi^{*}=\text{sign}(\mathcal{M}_{j3}) \frac{\sqrt{4\big(\kappa^2+\frac{2 \kappa^4}{\pi}}\big)}{2+\frac{4 \kappa^2}{\pi}}. \label{delta_star}
\end{eqnarray}

\vspace{0.2cm} 

\noindent So, we have $\rho^{*}=\psi^*/\sqrt{1-(\psi^*)^2}$ and plugging (\ref{delta_star}) in (\ref{scale_isolate}), we recover:

\vspace{-0.2cm}

\begin{eqnarray}
\varsigma^{*}=\sqrt{\frac{\mathcal{M}_{j2}}{\big(1-\frac{2}{\pi} \ (\psi^*)^2\big)}}.
\end{eqnarray}

\vspace{0.2cm}

\noindent Finally, the location parameter is given by:

\vspace{-0.4cm}

\begin{eqnarray}
\mu^*=\mathcal{M}_{j1}-\varsigma^* \sqrt{\frac{2}{\pi}}\ \psi^*.
\end{eqnarray}

\vspace{0.2cm}

\noindent The skew-normal fit to the conditional $\tilde{p}(v_j \vert \hat{\mathbf{v}}_{-j},\mathcal{D})$ is denoted by $\text{SN}_j(\mu^*, \varsigma^{*2},\rho^*)$ and can be used for the grid construction strategy.

\bibliographystyle{apa}
\bibliography{Bibliography_GAM}

\end{document}